\documentclass[showpacs,preprintnumbers,superscriptaddress,amsmath,amssymb]{revtex4-2}
\usepackage[colorlinks=true,bookmarks=false,citecolor=blue,urlcolor=blue]{hyperref}
\usepackage{amsmath,amssymb}
\usepackage{graphicx}
\usepackage{verbatim}
\usepackage{enumitem}
\usepackage{color}
\usepackage[english]{babel}
\usepackage{bm}
\usepackage{natbib}
\usepackage[symbol]{footmisc}
\usepackage{textcomp}

\usepackage{setspace}
\begin{document}
\title{Desktop laboratory of bound states in the continuum in metallic waveguide
with dielectric cavities}
\author{Evgeny Bulgakov}
\affiliation{Kirensky Institute of Physics Federal Research Center
KSC SB RAS 660036 Krasnoyarsk Russia}
\author{Artem Pilipchuk}
\affiliation{Kirensky Institute of Physics Federal Research Center
KSC SB RAS 660036 Krasnoyarsk Russia}
\author{Almas Sadreev}
\thanks{Corresponding Author}
\email{almas@tnp.krasn.ru} \affiliation{Kirensky Institute of
Physics Federal Research Center KSC SB RAS 660036 Krasnoyarsk
Russia}
\date{\today}

\begin{abstract}
We consider dielectric cavities whose radiation space is
restricted by two parallel metallic planes. The TM solutions of
the Maxwell equations of the system are equivalent to the
solutions of periodical arrays of dielectric cavities. The system
readily allows to achieve bound states in the continuum (BICs) of
any type including topological BICs as dependent on position and
orientation of the cavities relative to the planes. That extremely
facilitates experimental studies in comparison to infinite arrays
of the cavities. We show the effect of merging of topologically
protected BICs that pushes the square asymptotic of the $Q$-factor
into the power degree 4 or even 6.
\end{abstract}
\maketitle
\section{Introduction}
Since the famous paper by Gustav Mie \cite{Mie1908} engineering of
dielectric cavities in optics and photonics has been a
long-standing area of implementation various ideas and approaches
to enhance the quality factor $Q$ due to its paramount importance
in both applied and fundamental research. Conventionally, light
can be confined in closed or Hermitian system where access to
radiation channels is prohibited due to, for example, metallic
covering or embedding the resonant frequencies of cavity into band
gap of photonic crystal \cite{Joan}. However that way conflicts
with necessity of easy manipulation of light confined in the
cavity. On the other hand, although the compact dielectric
resonator is in air its $Q$-factor is very restricted due to
embedding into the radiation continuum whose spectrum is given by
light cone $\omega=ck$ with no cutoff. In principle the problem
can be solved cardinally if to address to infinitely long
periodical structures, for example, gratings. The periodicity
quantizes the directions of radiation leakage by means of
diffraction orders and brings cutoffs. As a result the periodical
structures support bound states in the radiation continuum (BICs)
\cite{Vincent1979,Bonnet1994,Astratov1999,Tikhodeev2002,Shipman2005,
Marinica2008,Hsu2013,Bulgakov2014,Yang2014,Bykov2015,Hu2015,Hsu2016,
Gao2016,Sadrieva2017}. However, in practice an increasing
of the number of cavities in periodical arrays is limited by
material losses \cite{Sadrieva2019} and structural fluctuations
\cite{Maslova2021}. Moreover grating slabs yield isolated
dielectric cavities in compactness. In view of that breakthrough
in the engineering of dielectric cavities was achieved owing to
avoided crossing of resonances of single cavity
\cite{Rybin2017,Bogdanov2019,Koshelev2019,Odit2020,Huang2021} or
different cavities \cite{Bulgakov2021,Pichugin2021}.

\begin{figure}[ht!]
 \centering
\includegraphics[width=0.45\linewidth]{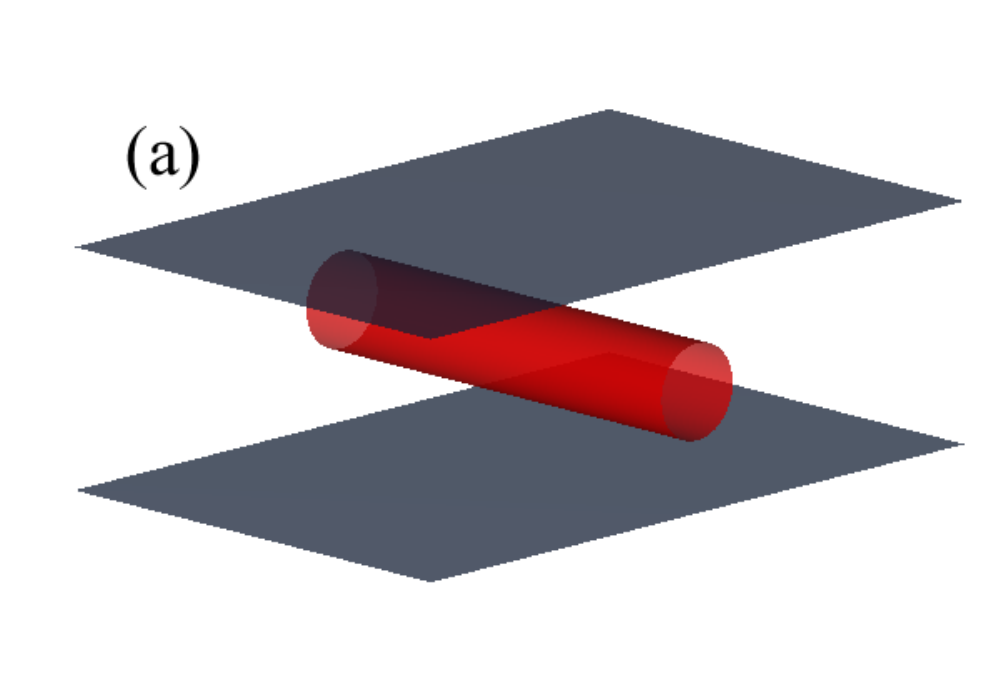}
\includegraphics[width=0.45\linewidth]{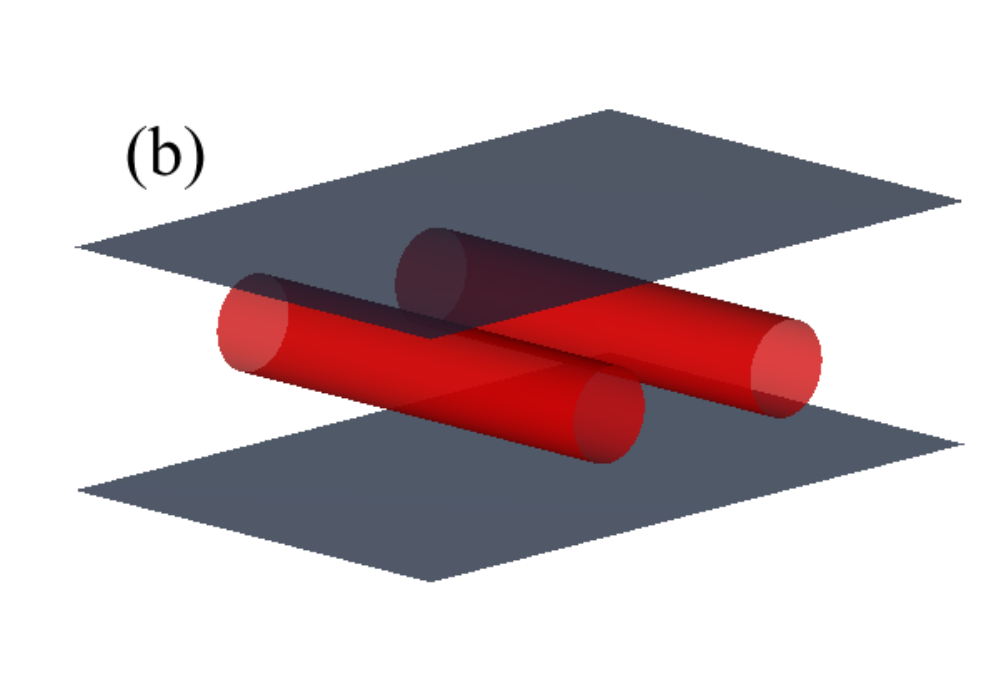}
\includegraphics[width=0.45\linewidth]{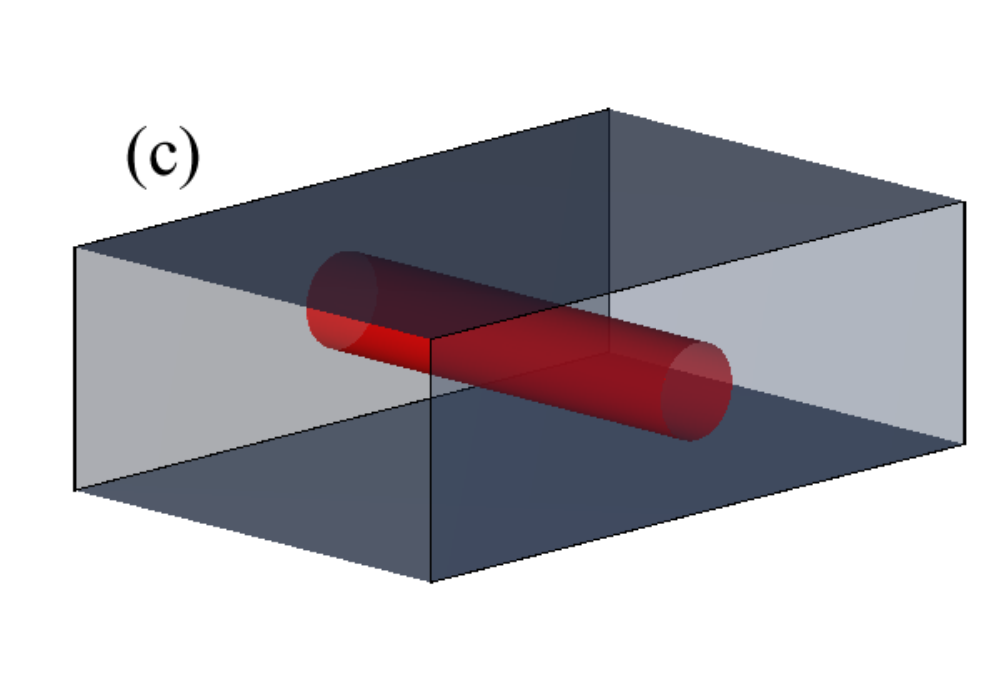}
\includegraphics[width=0.45\linewidth]{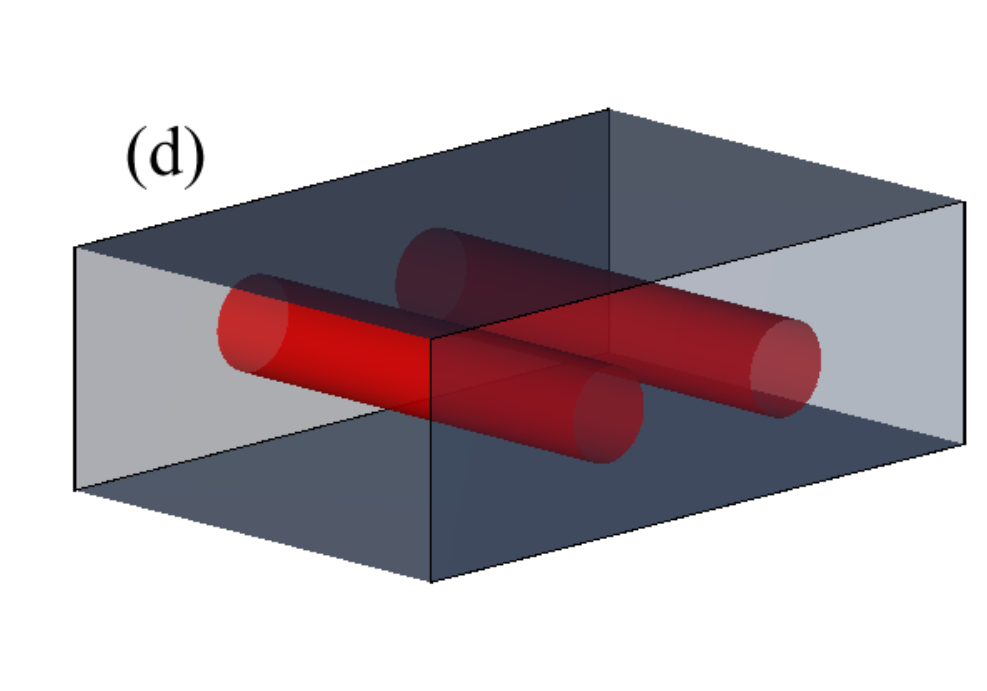}
\caption{ (Color online) Single (a) and two parallel (b) infinitely long  dielectric cylinders between two metallic planes. The same for cylinders of finite length inside waveguide of rectangular
cross-section (c) and (d).}
\label{fig1}
\end{figure}

In spite of reporting unprecedent values of the $Q$-factor in
these cavities that can not  achieve infinity because the isolated
cavity in air can not support the true  BICs
\cite{Colton,Silveirinha14}. In other words, although the
multipolar radiation with low orders of orbital momentum can be
suppressed due to avoided crossing of resonances the higher order
multipolar radiation still remains
\cite{Koshelev2018,Chen2019,Huang2021,Bulgakov2021}.  In the
present paper we propose compromise solution of the problem by
restriction of radiation space by  two parallel metallic planes
separated by a distance $d$. Then, for instance,  the single
dielectric cylinder inserted between the planes, as depicted in
Fig. \ref{fig1} (a), is equivalent to the infinite periodical
array of cylinders with the period $d$ in which BICs were
considered by many scholars
\cite{Shipman2005,Bulgakov2014,Bykov2015,Hu2015,
Yuan2017,B&M2017,Bulgakov2017c,Sadrieva2017,Yuan&Lu2018,Abujetas2019,Hu2020,Hu2020a}.
The equivalency follows from the Dirichlet boundary conditions at
perfectly conducting metal surface for the TM solutions of the
Maxwell equations. Respectively, two identical cavities between
the planes are equivalent  to two periodical arrays of rods which
support Fabry-Perot BICs \cite{Marinica2008,Ndangali2010}.

Moreover, the case of metallic waveguide of rectangular
cross-section allows to remove another typical theoretical
approximation of infinitely long cylinders rods as shown in Fig.
\ref{fig1} (a)-(d). Examples of equivalent arrays of dielectric
cavities are sketched in Fig. \ref{fig2}. Thus, the metallic
waveguide with one or two dielectric insets  is converting into
desktop laboratory of variety of BICs. While achievement of the
BICs in gratings often requires tuning of geometrical parameters,
the present case of desktop laboratory allows to achieve the BICs
by simple variation of the waveguide's width, position or
orientation of dielectric inset inside the waveguide. However, the
main advantage of this laboratory is related to material losses
and structural fluctuations which fast saturate growth of the
$Q$-factor with the number of cavities in periodic arrays
\cite{Sadrieva2019,Sidorenko2021,Maslova2021} that prevents to
clearly unveil the BICs.

The existence of symmetry protected (SP) BICs near rigid symmetric
obstacle cylinder placed symmetrically in between parallel walls
with Neumann or Dirichlet boundary conditions imposed upon them
was proven on papers  \cite{Evans1994,Evans1998,Newman2016}.
After, Linton {\it al} \cite{Linton2002} and Duan {\it et al}
\cite{Duan2007} have examined the cases of accidental BICs for
slender rods of arbitrary cross-section placed non-symmetrically
in waveguide. We develop these results for the case of dielectric
rods of circular and rectangular cross-section and find a
threshold for dimensions and permittivity of the rod below which
the BICs do not exist.
Once the rod is shifted  relative to center line we obtain
equivalent dimerised chain of rods as shown in Fig. \ref{fig2}
(c). There are only a few reports of BICs in dimerised arrays of
dielectric cavities  \cite{Overvig2018,Song2019}. Moreover, the
rectangular rod inserted between metallic planes brings new
parameter to vary, the angle of orientation of the rod as shown in
Fig. \ref{fig2} (d). That, in turn, opens a way for
realization of topologically protected BICs with winding
numbers $m=\pm 1$ in two-parametric space of angle and frequency
or angle and distance between rods.

\section{Cylinder between two metallic planes}
\begin{figure}[ht!]
 \centering
\includegraphics[width=0.4\linewidth]{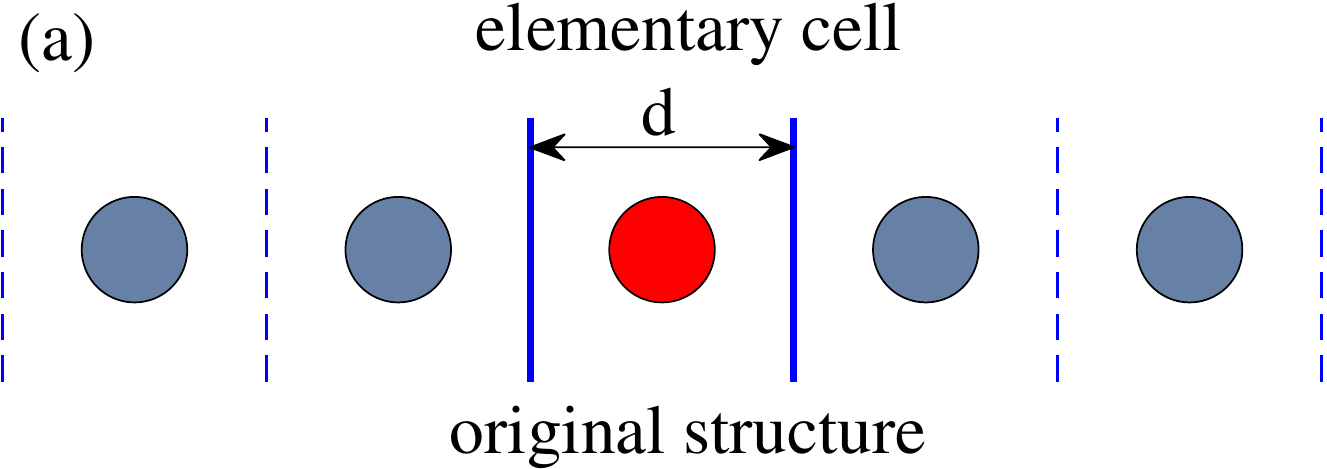}\hspace{0.05\linewidth}
\includegraphics[width=0.4\linewidth]{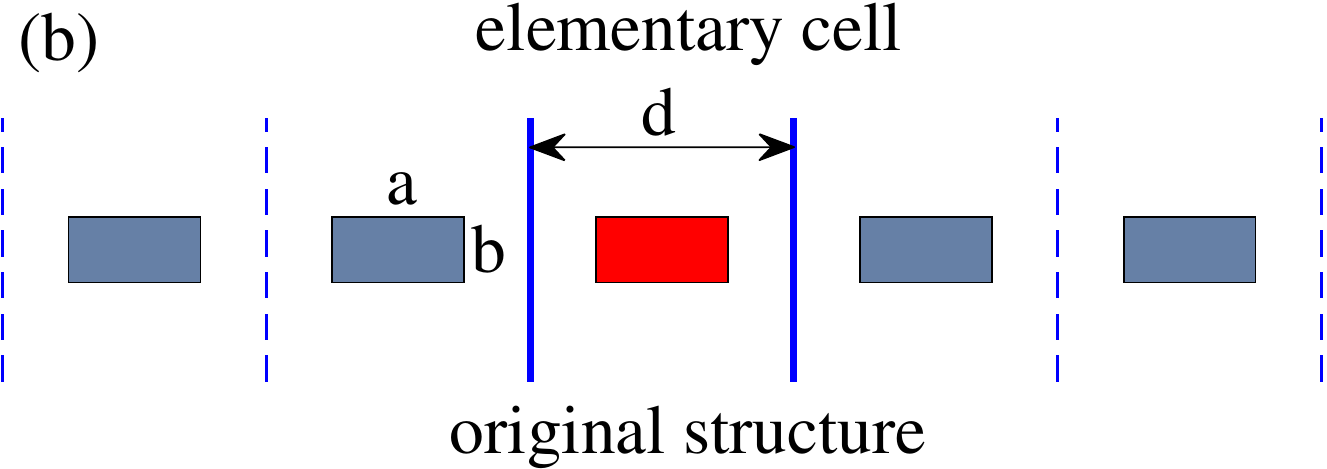}\vspace{1cm}
\includegraphics[width=0.4\linewidth]{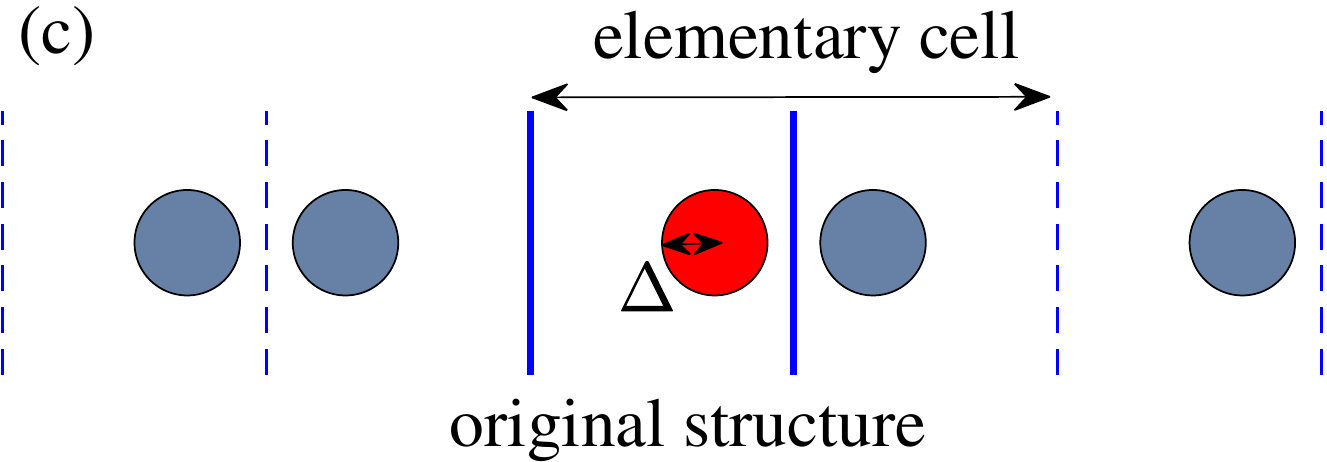}\hspace{0.05\linewidth}
\includegraphics[width=0.4\linewidth]{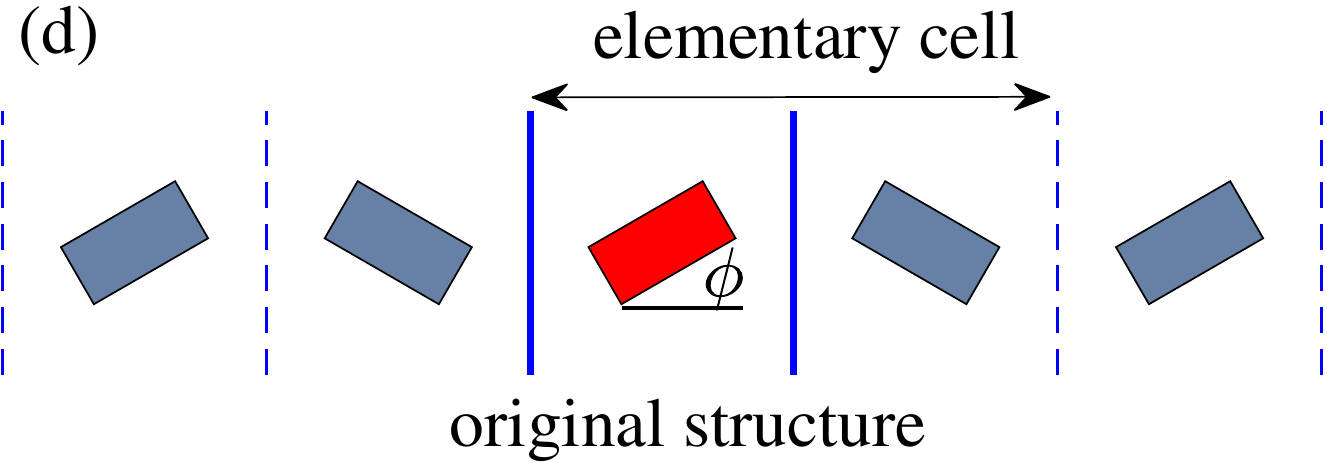}\vspace{1cm}
\includegraphics[width=0.4\linewidth]{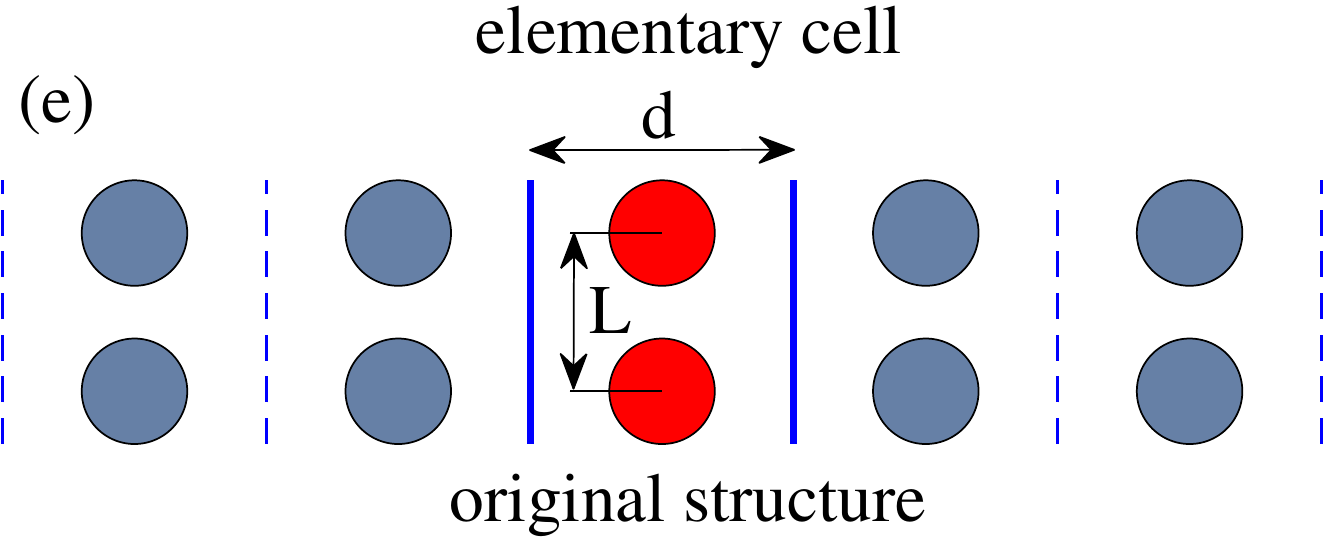}\hspace{0.05\linewidth}
\includegraphics[width=0.4\linewidth]{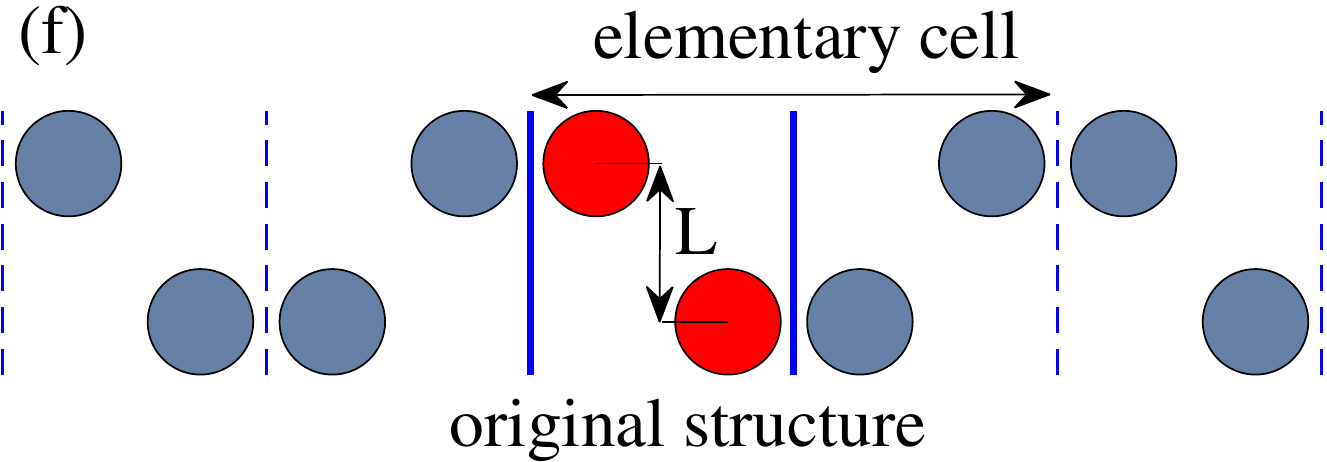}\vspace{1cm}
\includegraphics[width=0.4\linewidth]{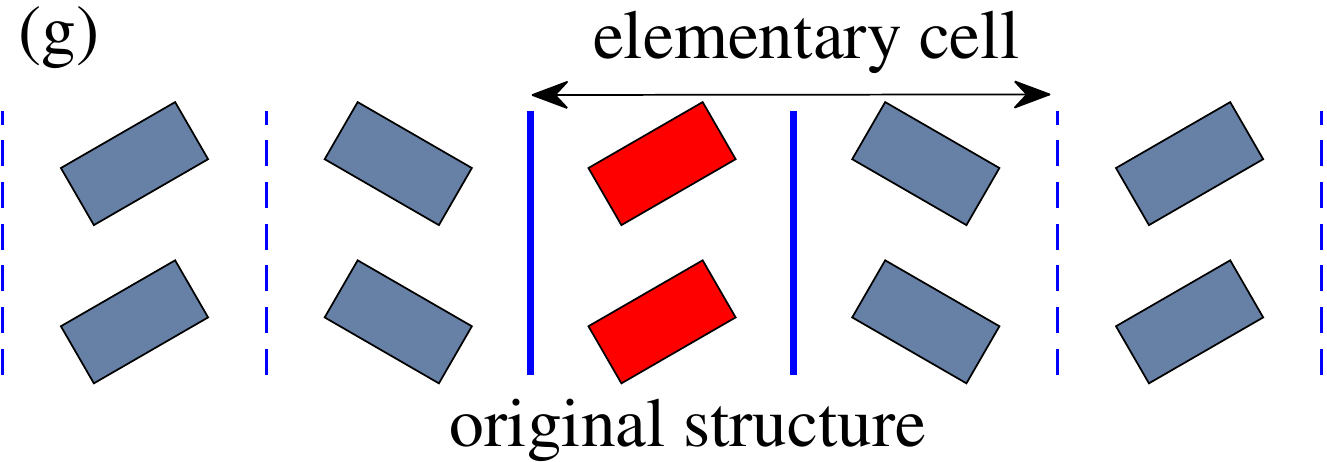}\hspace{0.05\linewidth}
\includegraphics[width=0.4\linewidth]{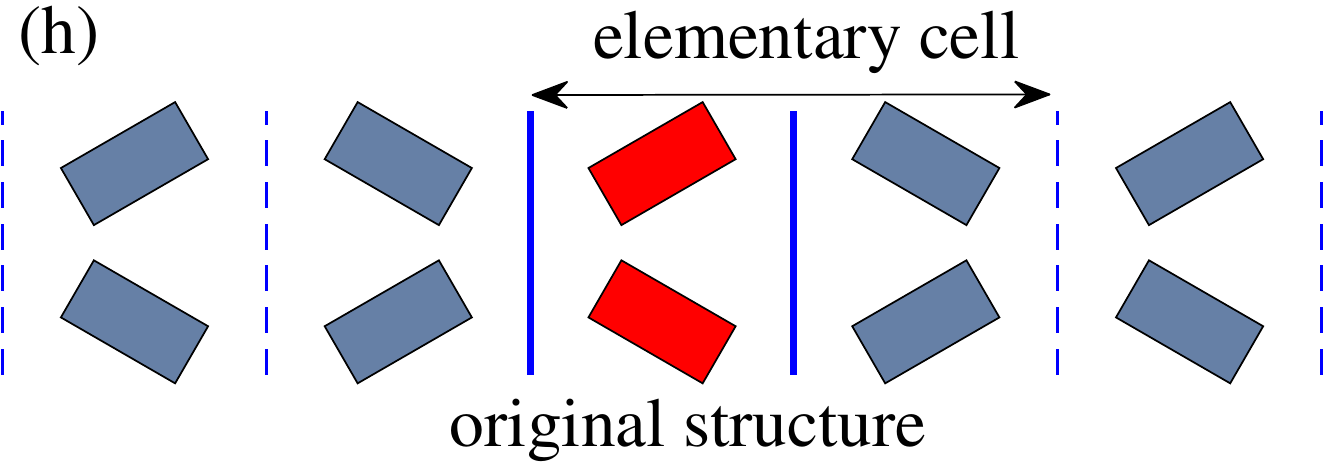}
\caption{ (Color online) Periodical arrays of dielectric rods whose solutions at
the $\Gamma$-point are equivalent to the solutions of Maxwell
equations of rods placed between two parallel metallic planes shown by
solid thick lines. The original rod is shown by red while its
images at metallic planes are shown by gray. Cylindrical (a) and
rectangular (b) rods positioned symmetrically between planes are
equivalent to the periodical array of rods with period $d$.
Cylindrical rod shifted by distance $y_0$ from center line (c) and
rectangular rod rotated by the angle $\phi$ makes the system
equivalent to dimerised chain. (e) Two rods positioned symmetrically
make the system equivalent to double array of rods which can
support FPR BICs. Two circular rods, positioned non-symmetrically (f)
or two rotated rectangular rods (g) and (h) are equivalent to two dimerised chains. }
\label{fig2}
\end{figure}

We start consideration with single dielectric cylinder of the
radius $R$ inserted parallel to metallic planes as depicted in
Fig. \ref{fig1} (a). In what follows all quantities are measured
in terms of the distance between plates $d$ with $x$-axis is
directed along the waveguide and $z$-axis is directed along the
rod. Because of boundary conditions for the electric field $E_z(x
= 0, d) = 0$ the solutions of the Maxwell equations of the system
are equivalent to the solutions in periodical infinite array of
rods at the $\Gamma$- or $X$-point as sketched in Fig. \ref{fig2} (a).

There are two distinct cases for these solutions. The case of
dielectric inset symmetrically disposed between two parallel
metallic planes is equivalent to periodical array of rods with the
period $d$ as shown in Fig. \ref{fig2} (a) and (b). Fig.
\ref{fig2} (c) presents the case of cylinder shifted from the
center line of waveguide by distance $\Delta$. Then the TM
solutions of the system coincide with the solutions of the binary
array of cylinders with double period $2d$. The equivalence of the
solutions allows us to use well known analytical approaches
developed for periodical arrays of dielectric resonators
\cite{Yasumoto,Kushta2000}. We complement these approaches by
COMSOL MultiPhysics numerical calculations for the rods of
circular and rectangular cross-sections.

\begin{figure}[ht!]
\centering
\includegraphics[width=0.5\linewidth]{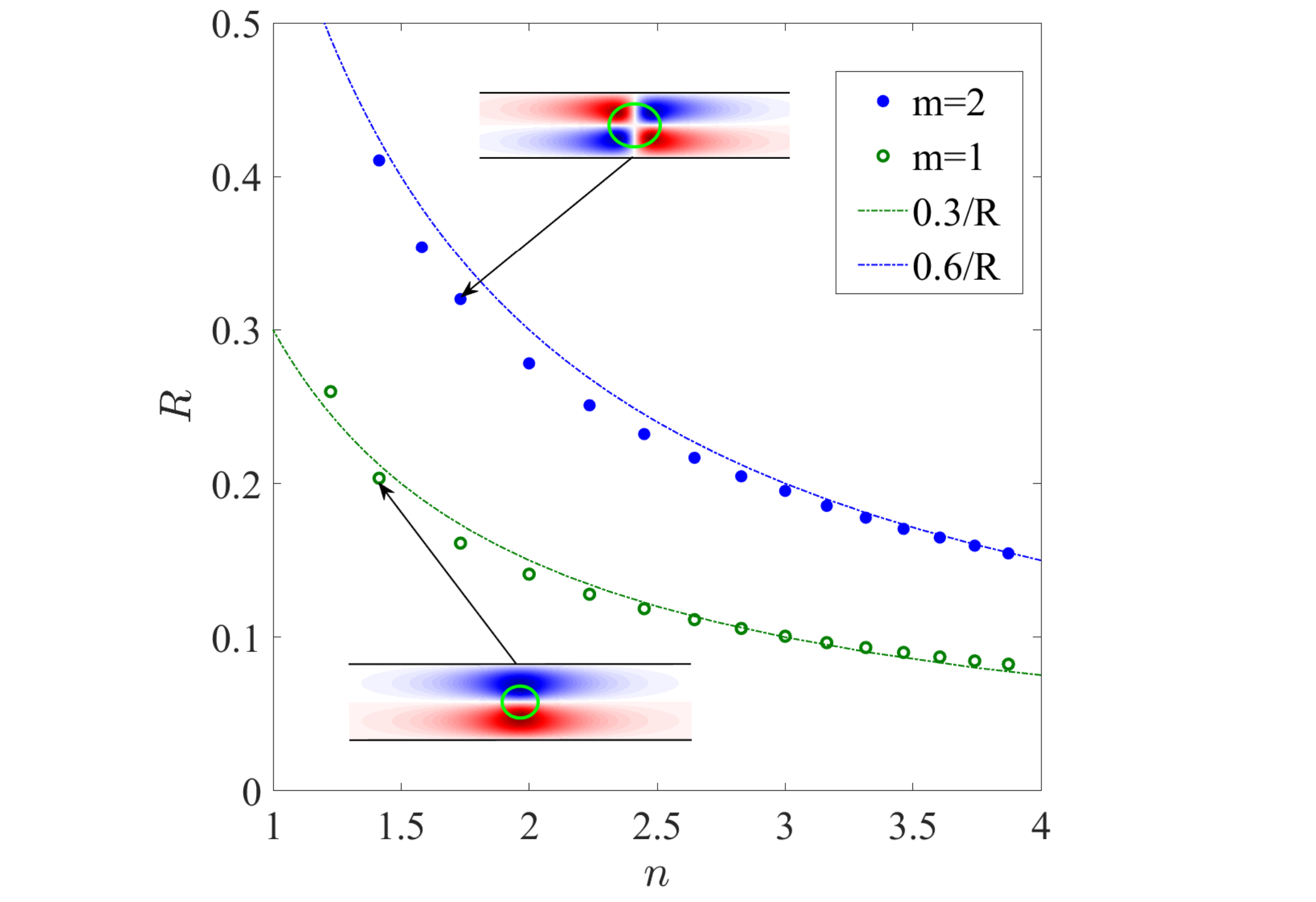}
\caption{ (Color online) Curves of existence of the SP BICs vs
refractive index and radius of circular rod. Insets show
patterns of SP BICs (Electric field $E_z$)}. \label{fig3}
\end{figure}

\begin{figure}[ht!]
\centering
\includegraphics[width=0.5\linewidth]{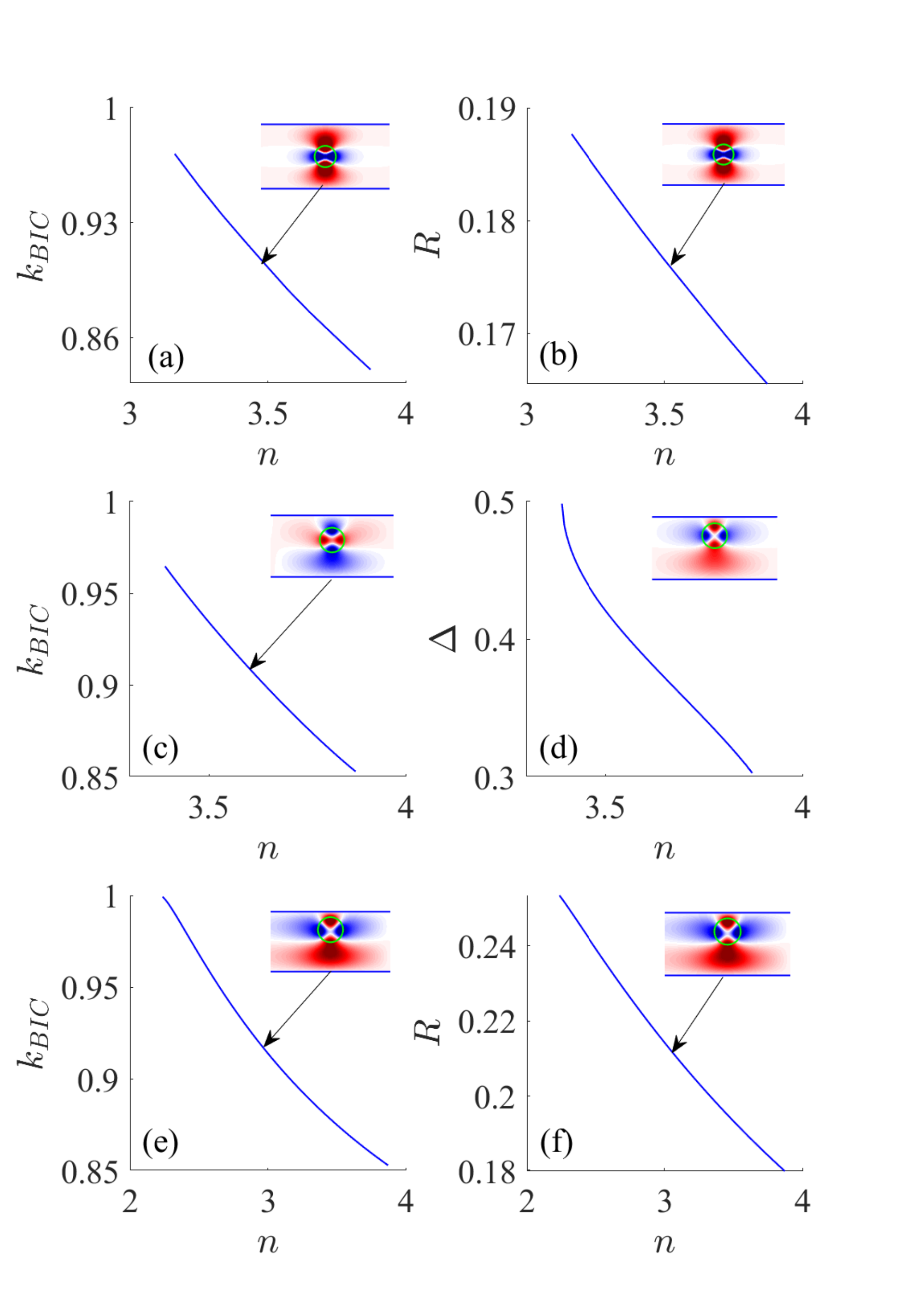}
\caption{ (Color online) Curves of existence of accidental BICs in two-parametric space of refractive index of the rod and
(a) frequency, (b) radius of the rod for $\Delta = 0$;
(c) frequency, (d) displacement of the rod for $R = 0.18$;
(e) frequency, (f) radius of the rod for $\Delta = 0.3$.
Insets show the accidental BICs for certain parameters. }
\label{fig4}
\end{figure}

In Fig. \ref{fig3} we show typical examples of the SP BICs with
azimuthal indices $m = 1$ and $m = 2$ for the case of symmetrical
position of dielectric cylinder with the refractive index $n$
(Fig. \ref{fig2} (a)). It is clear that in order the dielectric
rod could trap the EM wave with definite wavelength
$\frac{2\pi}{k}$ its radius $R$ has to be comparable with the
characteristic wavelength inside the rod $\frac{2\pi}{nk}$.
Therefore the curve of existence of the SP BICs can be evaluated
as $Rnk_{BIC}(n,R)\approx 1$. Numerical behavior and comparison
with this evaluated formula is shown in Fig. \ref{fig3}. Since the
diameter of rod can not exceed distance between mirrors we obtain
that BICs can not exist for $nk_{BIC}\geq 2$.
For non-symmetric position of the rod inside the waveguide the SP BICs
transform to accidental BICs at tuned rod's radius or refractive
index. In the equivalent system of binary periodical array of rods
(see Fig. \ref{fig2} (c)) these BICs correspond to the BICs at
$\Gamma$-point. Phase diagrams of existence of the accidental BICs
are plotted in Fig. \ref{fig4}.

\section{Rectangular rod between two metallic planes}

\begin{figure}[ht!]
\centering
\includegraphics[width=0.5\linewidth]{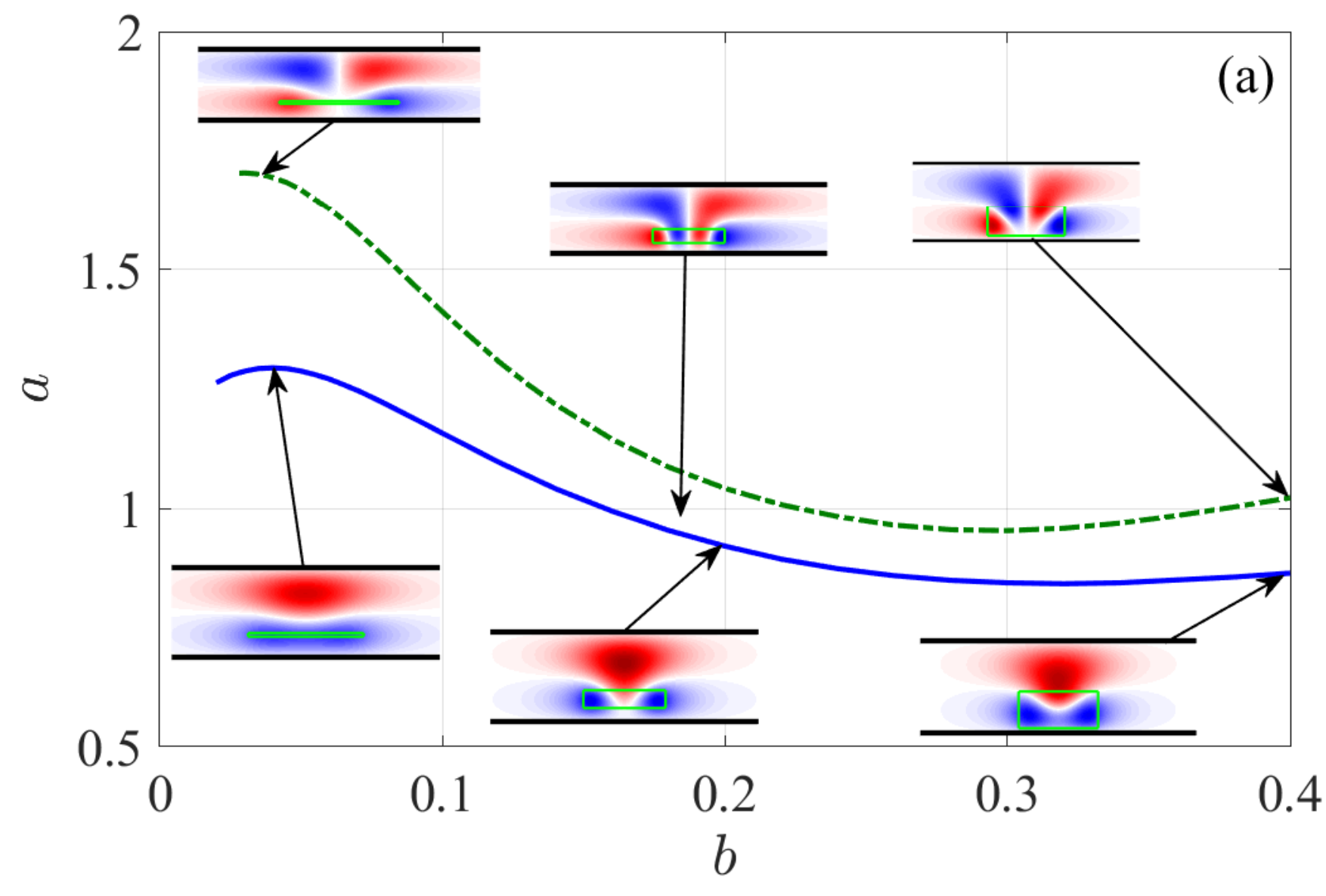}
\includegraphics[width=0.5\linewidth]{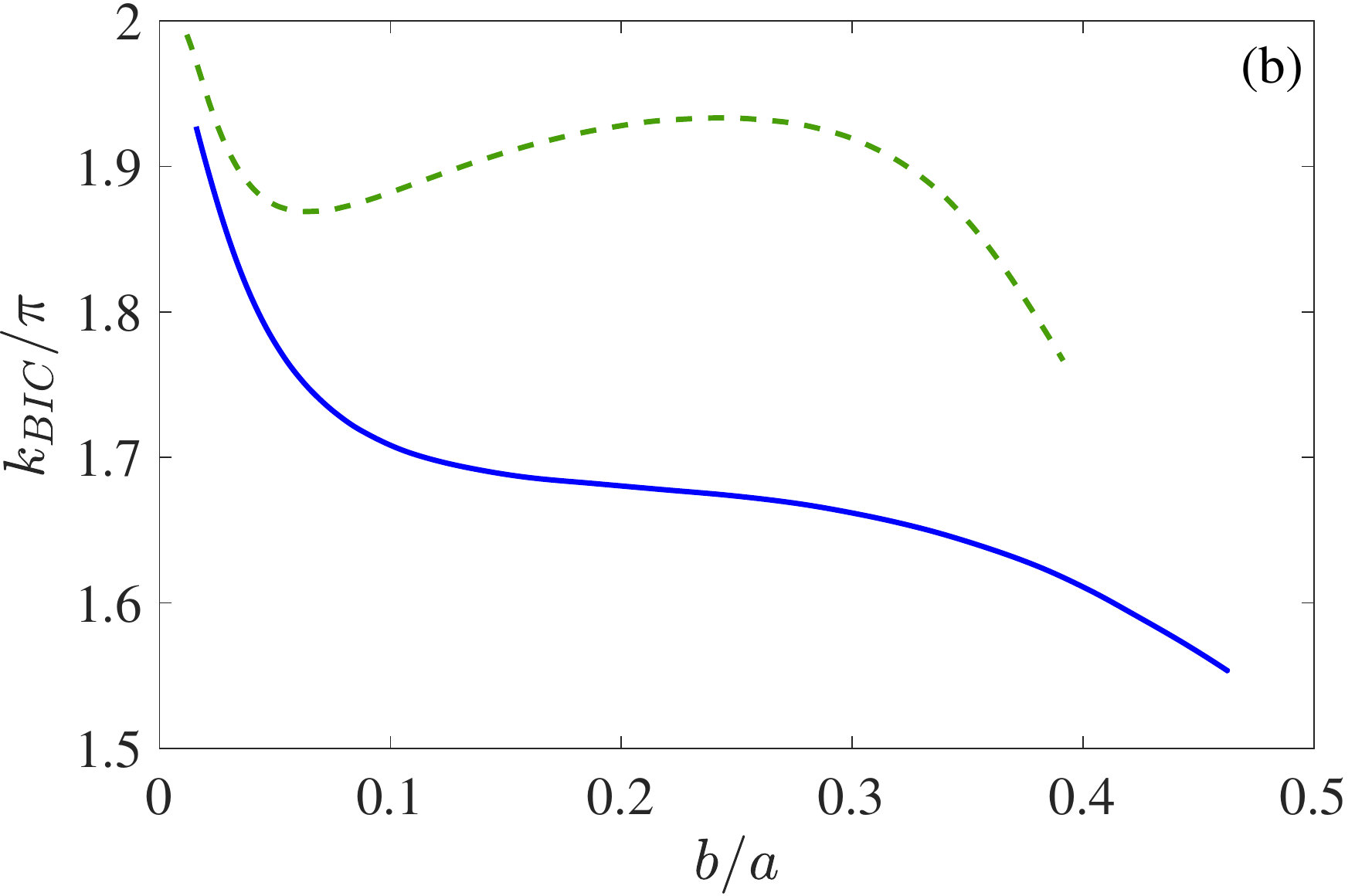}
\caption{ (Color online) Curves of existence of the accidental BICs. (a) As
dependent on cross-section of rectangular quartz rod $a\times b$
at $\Delta=0.25$ with refractive index $n=2.05$. (b) The frequency
of BICs vs aspect ratio of rectangular rod.}
\label{fig5}
\end{figure}

First of all, the rectangular rod is interesting by that Linton
{\it et al} \cite{Linton2002} have shown existence
of accidental BICs under assumption that the aspect ratio $a/b$
is sufficiently large and the rod is metallic. Moreover,
rectangular rod brings new parameter to vary, rotation angle
relative to the frame of metallic waveguide as shown in Fig.
\ref{fig2} (d). Fig. \ref{fig5} (a) shows curves of accidental
BICs versus dimensions of rectangular quartz rod $a \times b$.
Insets show evolution of two accidental BIC modes (electric field
$E_z$ directed along the rod). In fact, there are more curves which
differ by number of nodal lines cross to the waveguide. TM
propagating channels are given by simple formula
$k^2 = k_x^2 + \pi^2 p^2, p=1, 2, 3, \ldots$ where $k_x$ is the wave
vector of TM waves along the waveguide. We consider only the BICs
embedded into the continuum of the first propagating channel of the
waveguide with $p=1$ and frequencies below the cutoff of the
second propagating channel $\pi < k < 2\pi$.

First, one can see that the curves of BICs follow to analysis
of Linton {\it et al} obtained by different mathematical
techniques \cite{Linton2002} for the case of Dirichlet BC at the
walls of rectangular rod, i.e., metallic rod. These accidental
BICs shown in Fig. \ref{fig5} (a) have clear physical origin. Far
from the rod accidental BICs follow to the evanescent mode $p = 2$
which is orthogonal to the first propagating channel $p = 1$ and
therefore can not go out. The dielectric rod perturbs the
evanescent mode and the perturbation strength depends on size and
refractive index of the rod. The more the index and size the more
perturbation. In spite of difference between the metallic and
dielectric rods the accidental BIC exists at $b \rightarrow 0$.
However, BIC's frequency is limited by the cutoff $2\pi$ of the
second channel and the localization range diverges for
$k \rightarrow 2\pi$ as insets in Fig. \ref{fig5} (b) show.

\section{Fabry-Perot BICs: two rods inside waveguide}
\begin{figure}[ht!]
\centering
\includegraphics[width=0.75\linewidth]{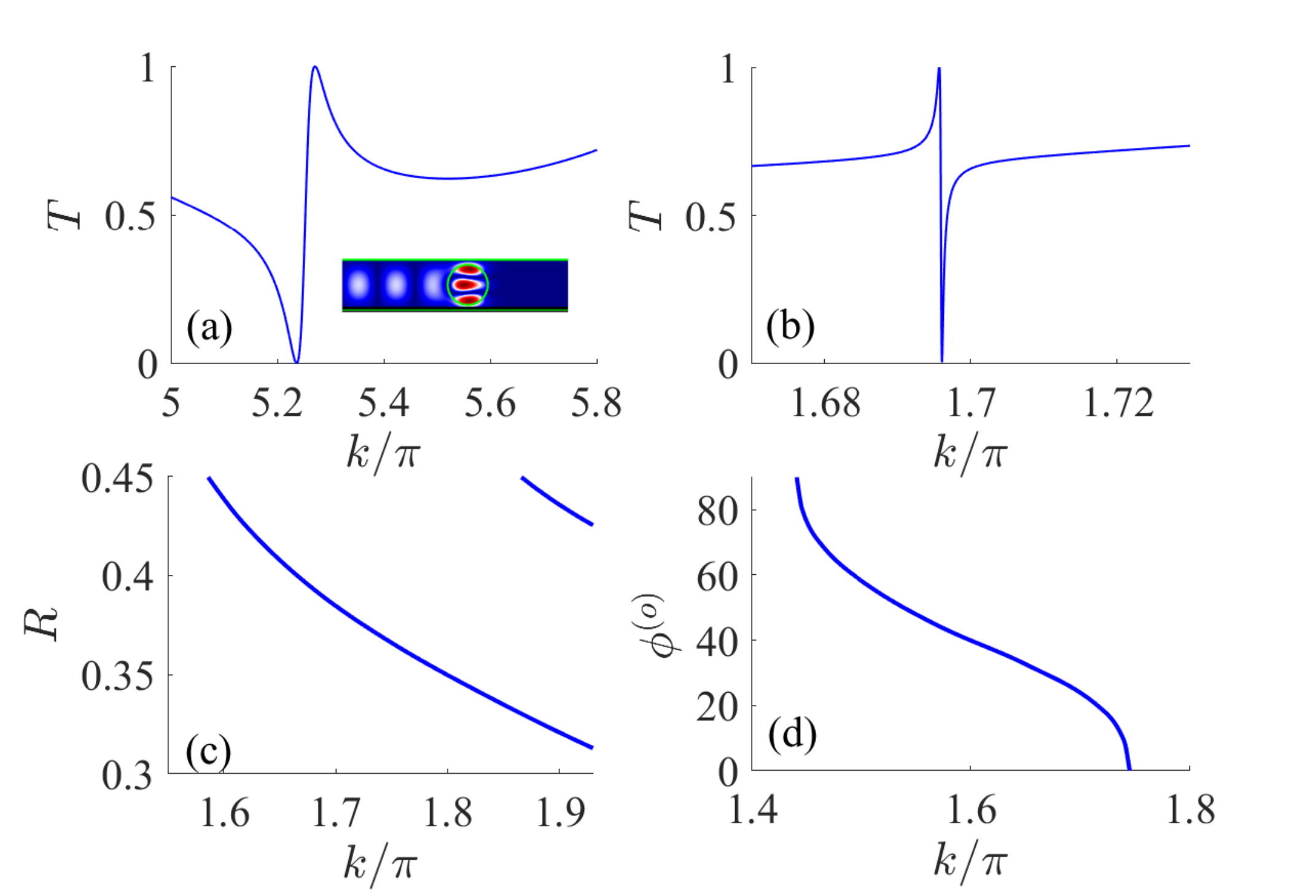}
\caption{(Color online)
Transmittance vs frequency (upper panels) for
(a) cylindrical rod with $R = 0.4, \Delta = 0$,
(b) rectangular rod with cross-section $0.6 \times 0.3$ and $\Delta = 0.01$.
Transmittance zeros (bottom panels) vs frequency of incident wave and
(c) radius of cylindrical rod,
(d) orientation angle of rectangular rod for $\Delta = 0$.
The refractive index of the rods is $n = 2.05$ in all cases. }
\label{fig6}
\end{figure}

Two identical circular rods inserted symmetrically inside the
waveguide as shown in Fig. \ref{fig1} (e) make the system
equivalent to periodic double arrays of subwavelength dielectric
cylinders. Such arrays were studied by Ndangali and Shabanov with
analytic TM solutions for BICs in the limit of thin cylinders
\cite{Ndangali2010}. Underlying physical mechanism for BICs is the
Fabry-Perot one \cite{Sadreev2021} in which each array perfectly
reflects electromagnetic waves at definite frequency $k_c$. Then,
two arrays serve as ideal mirrors which are able to capture
electromagnetic wave with frequency $k_c$ at discrete distances
roughly equal to integer number of half wavelength $\pi/k_c$. Fig.
\ref{fig6} demonstrates the effect of total reflection of TM waves
by (a) cylindrical and (b) rectangular rods positioned
symmetrically inside the waveguide. The effect of total reflection
by rod inserted into waveguide exists irrespective to position of
the rod inside the waveguide. In Fig. \ref{fig6} (b) one can see the
well known effect of collapse of Fano resonance for
$\phi\rightarrow 0$ and $\Delta=0$ at which the scattering
function tends to the SP BIC \cite{Kim1999}.
Fig. \ref{fig6} (c) shows that total reflection ($T = 0$) is achieved
owing to variation of cylindrical rod radius. However,
the variation of the rod size is difficult in the experiment. Fig.
\ref{fig6} (d) demonstrates that the problem can be easily solved
by orientation of rectangular rod relative to waveguide.
\begin{figure}[ht!]
\centering
\includegraphics[width=0.4\linewidth]{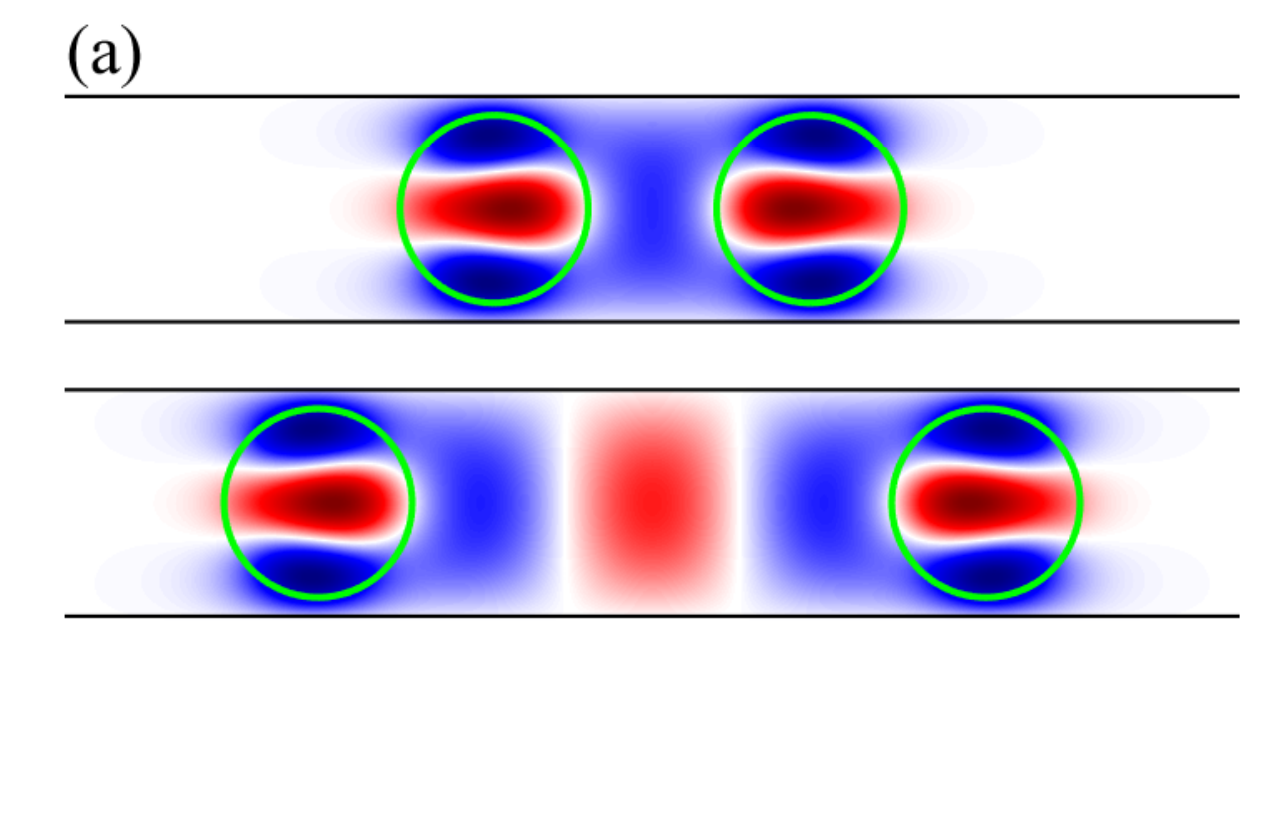}
\includegraphics[width=0.4\linewidth]{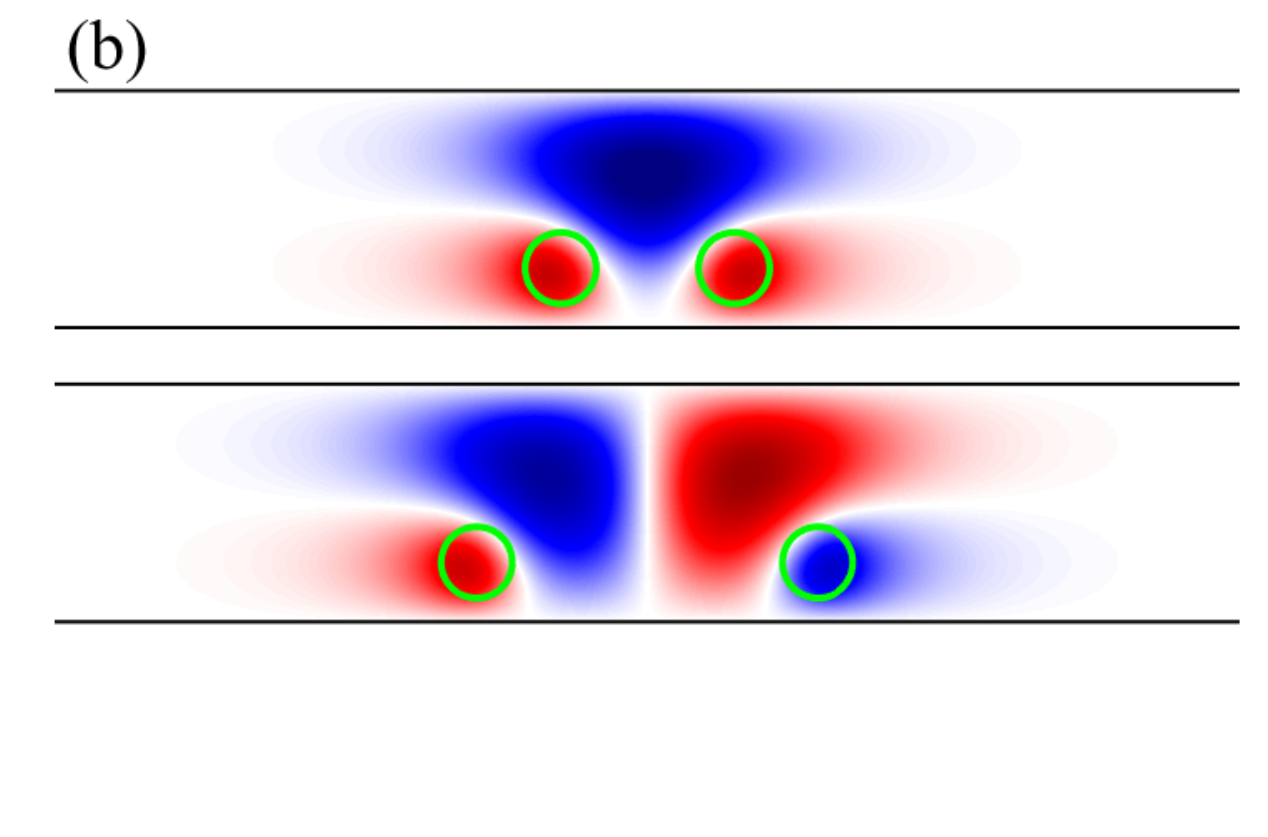}
\caption{(Color online) Patterns of Fabry-Perot BICs for the case
of two cylindrical rods (shown by green circles) positioned (a)
symmetrically ($R=0.4, n = 2.05$) and (b) shifted cross to the
waveguide inside the waveguide by $\Delta=0.25, R=0.15$.}
\label{fig7}
\end{figure}

\begin{figure}[ht!]
\centering
\includegraphics[width=0.5\linewidth]{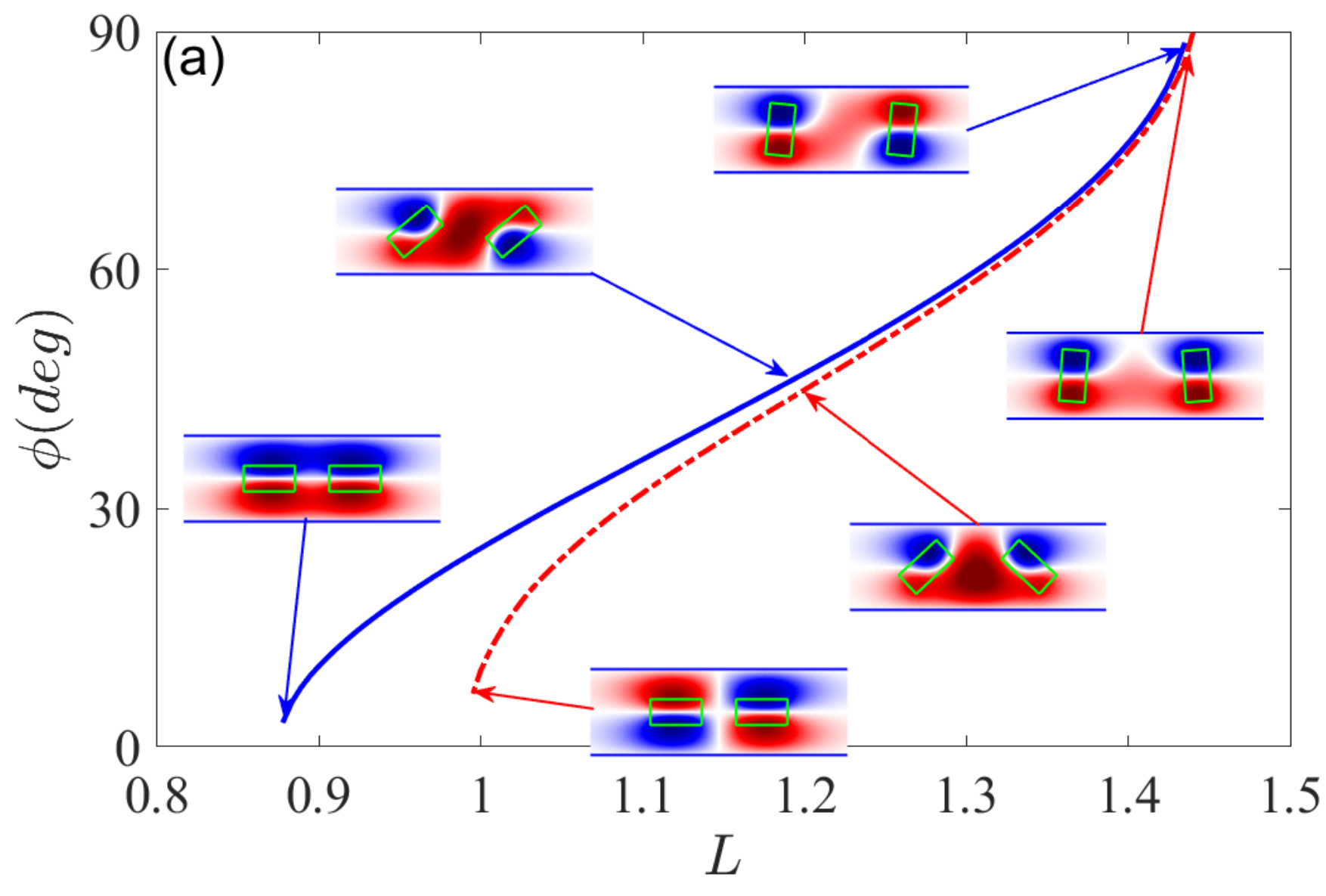}
\includegraphics[width=0.5\linewidth]{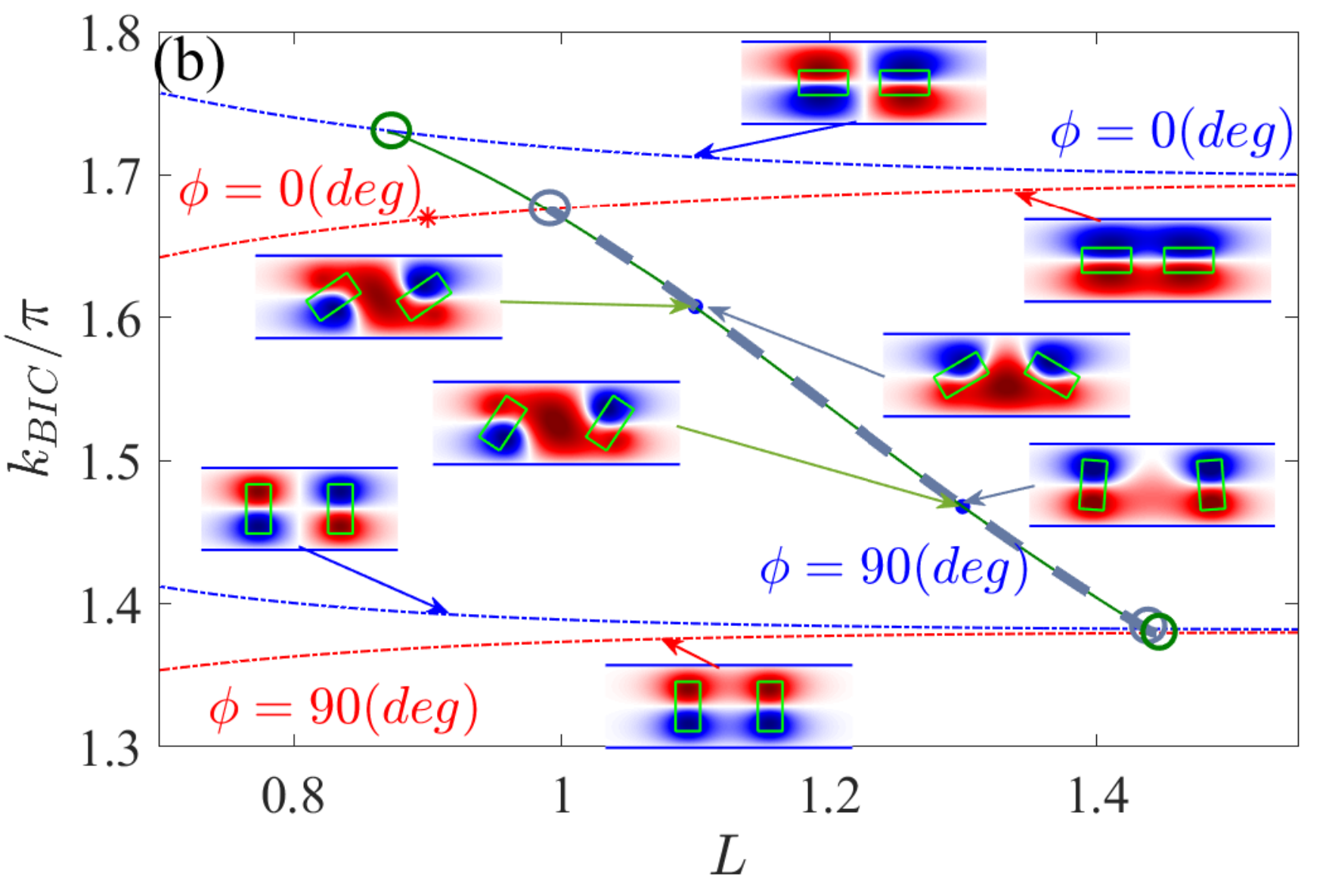}
\caption{ (Color online) Curves of existence of topologically protected BICs
for two rectangular rods inside the waveguide.
(a) Vs distance between rectangular rods and angle of rotation,
(b) vs frequency and distance between insets at $\Delta = 0$.
Points of merging are marked by open circles.}
\label{fig8}
\end{figure}

In Fig. \ref{fig7} (a) we show results of numerical
calculations for the case of two circular quartz rods inserted
symmetrically (see Fig. \ref{fig2} (c)) with patterns of
Fabry-Perot BICs for different distances between cylindrical rods.
That case is equivalent to the case of two periodic arrays
considered in Refs.
\cite{Liu2009,Ndangali2010}. It is interesting that similar type of BICs exist even for non symmetrical position of circular rods
an one can see from Fig. \ref{fig7} (b).
 In Fig. \ref{fig8} we show curves of
the Fabry-Perot BICs in two-parametric space of the
rotation angle of rods $\phi$ and distance between them $L$ for two distinct
cases of rods rotation: in-phase and anti-phase.

\begin{figure}[ht!]
\centering
\includegraphics[width=0.5\linewidth]{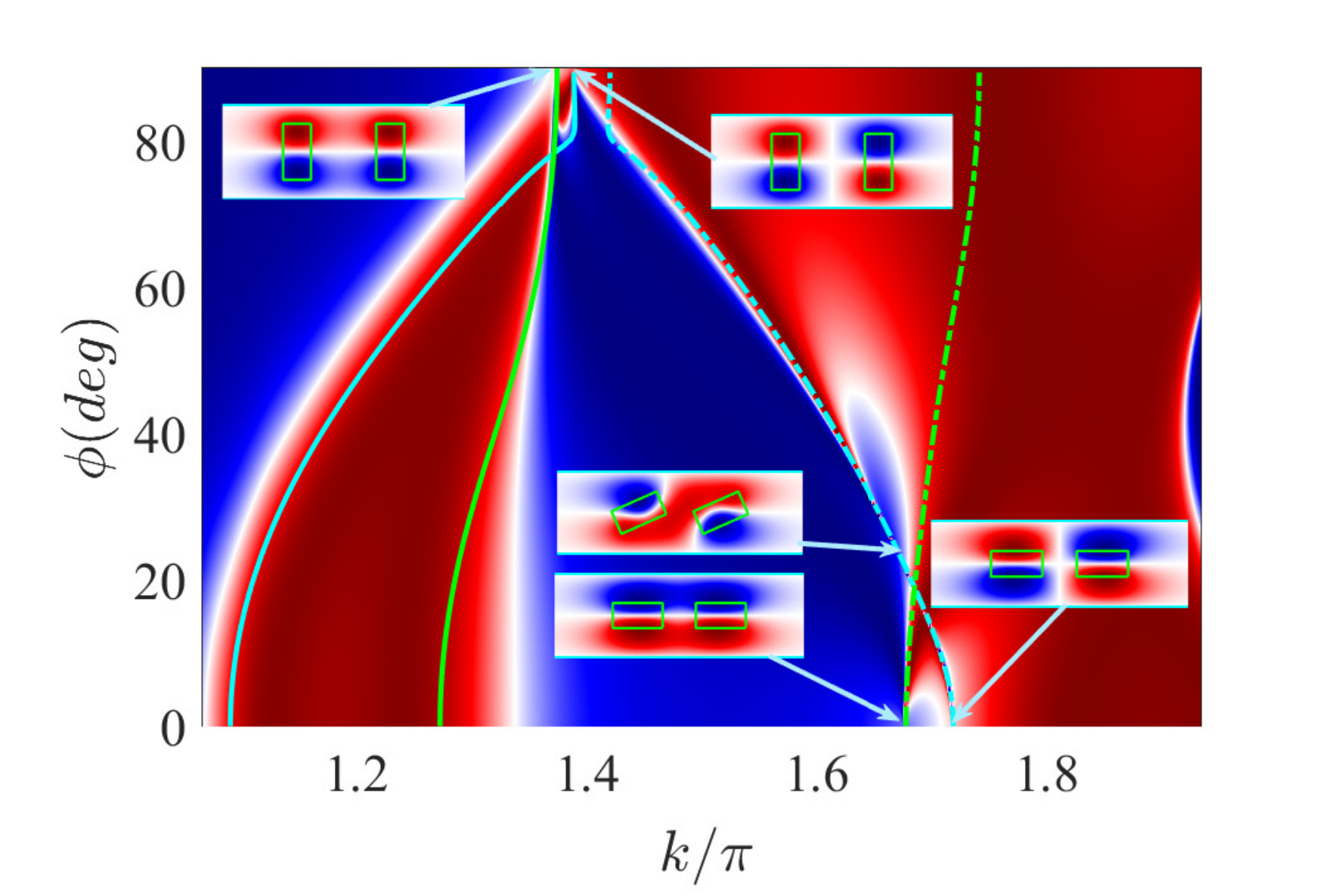}
\caption{ (Color online) Transmittance of electromagnetic waves
over the waveguide vs frequency and rotation angle of rods as
shown in insets. Lines show resonant frequencies as function of
the rotation angle $\phi$. The parameters are $a=0.6, b=0.3 L=1$. }
\label{fig9}
\end{figure}

One of the ways to experimentally confirm BICs is observation of
singularities in the wave transmission in waveguide. At the BIC
point the total reflection coalesces with the full transmission
\cite{SBR} which can be defined as collapse of Fano resonance
\cite{Kim1999}. In Fig. \ref{fig9} we present typical examples of
such singular points in parametric space of incident
light frequency and orientation angle of rectangular rods. Moreover, one
can see that the transmission peaks follow the resonant
frequencies of the system shown by solid lines in Fig. \ref{fig9}.
There are two equivalent cases $\phi=0$ and $\phi=\pi/2$ which
were considered in Refs. \cite{Marinica2008,Chen2022}. In both
cases we observe two solutions for BICs whose frequencies are
considered as splitting due to interaction of rods.
As a result we obtain the symmetric and
anti symmetric hybridized solutions shown in insets of Fig.
\ref{fig9}. If the rods were in air the splitting of frequencies
would decrease  as $1/L^2$ \cite{Bulgakov2019}. However a presence
of parallel metal planes cardinally changes the interaction
between two rods to cancel $L$-dependence.

\section{Topologically protected BICs merge into SP or accidental BICs}
\begin{figure}[ht!]
    \centering
    \includegraphics[width=0.5\linewidth]{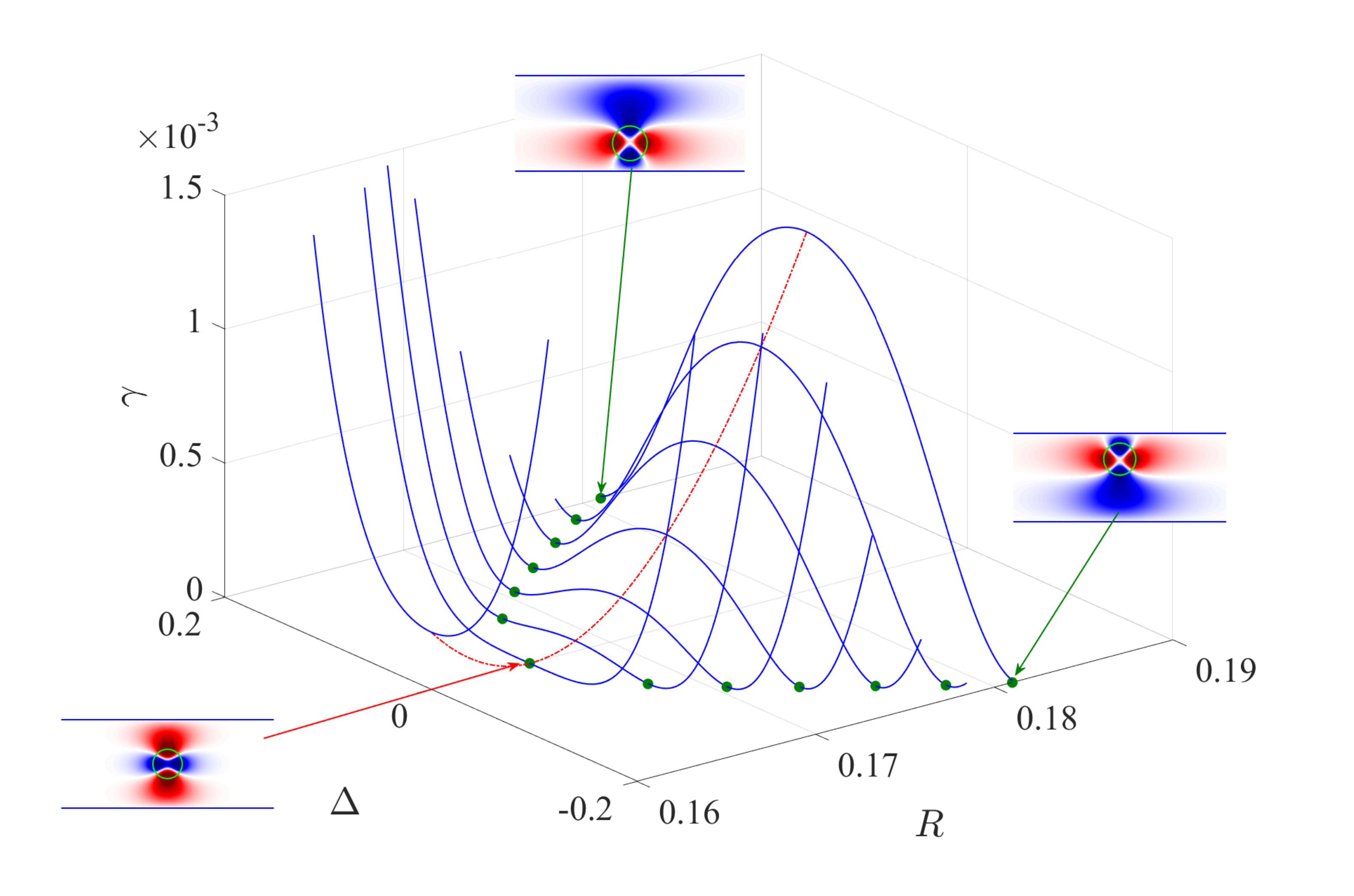}
    \caption{(Color online) The phase of function $\theta=Arg\left(\frac{1}{E_z}\right)$ vs frequency and the rotation angle of dielectric rectangular rod $0.6\times 0.3$
        for the case of inphase rotation for the parameters highlighted by open circle in Fig. \ref{fig9}. }
    \label{fig10}
\end{figure}

\begin{figure}[ht!]
\centering
\includegraphics[width=0.5\linewidth]{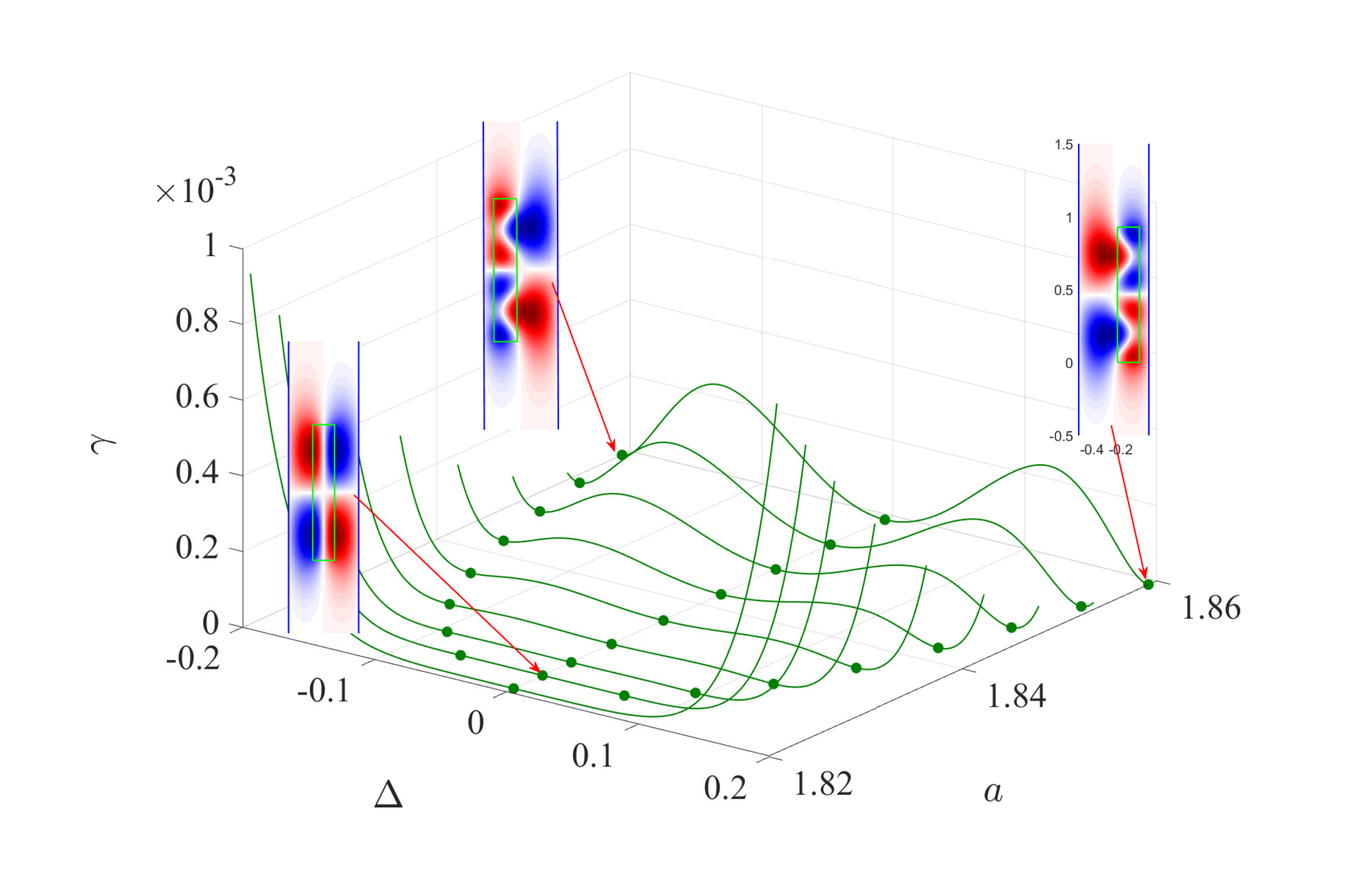}
\caption{ (Color online) Half width of resonances vs shift of cylindrical rod
relative to center of waveguide and radius of rod for
$n = 3.87$. Insets show patterns of BICs ($E_z$ of electric
field). } \label{fig11}
\end{figure}

Topologically protected (TP) BICs were reported by many scholars
\cite{B&MPRL,B&M2017,Mukherjee2018,Liu2019,Jin2019,Yoda2020,Hwang2021}
which originate from the merging of several BICs in the momentum
space. Very recently Huang {\it et al} \cite{Huang2022} have
demonstrated TP BICs in coupled acoustic resonators which arise
from the merging of BICs in parametric space of frequency and
coupling strength. The importance of the TP BICs is that they are
robust to the fabrication imperfection, and that the degree of
enhancement of the $Q$-factor of quasi BICs changes from standard
quadratic to the fourth or even to the sixth degree. Here we
demonstrate the cases of merging of two accidental BICs with
winding numbers  $m=\pm 1$ into one non-robust accidental BIC with
$m=0$. The phase singularities arise if some complex function
$\Psi(k,\phi)=u(k,\phi)+iv(k,\phi)=|\Psi(k,\phi)|exp(i\theta(k,\phi))$
has nodal point in some two-dimensional parametric space, for
example, frequency and angle of orientation $(k,\phi)$ or
$(k,\Delta)$. Then the winding number of the singularity is given
by
\begin{equation}\label{mm}
    m={\rm
    sgn}\left(\frac{\partial{u}}{\partial{k}}\frac{\partial{v}}{\partial{\phi}}-
    \frac{\partial{u}}{\partial{\phi}}\frac{\partial{v}}{\partial{k}}\right)
\end{equation}
or \begin{equation}\label{mmm}
    m=\oint d\overrightarrow{l}\nabla \theta.
\end{equation}
As the parametric space we have chosen the frequency $k$ and the
angle of rotation $\phi$ while for the function $\Psi$ we have
chosen $\frac{1}{E_z}(x,y)$ at some fixed space point $(x,y) $ where $E_z$ is the z-th component of electric
field for the TM solution of the Maxwell equations. Fig. \ref{fig10} shows as the phase of this
function in an anticlockwise sense, i.e., gives us $m=1$.
The merging is shown in Fig. \ref{fig11} in which
evolution of TP BIC with $m=\pm 1$ in two-parametric space of
radius and shift of dielectric cylinder is plotted by solid green
line while the accidental BIC with $m=0$ occurs only at $R=0.1656$ and
$\Delta=0$.
\begin{figure}[ht!]
\centering
\includegraphics[width=0.5\linewidth]{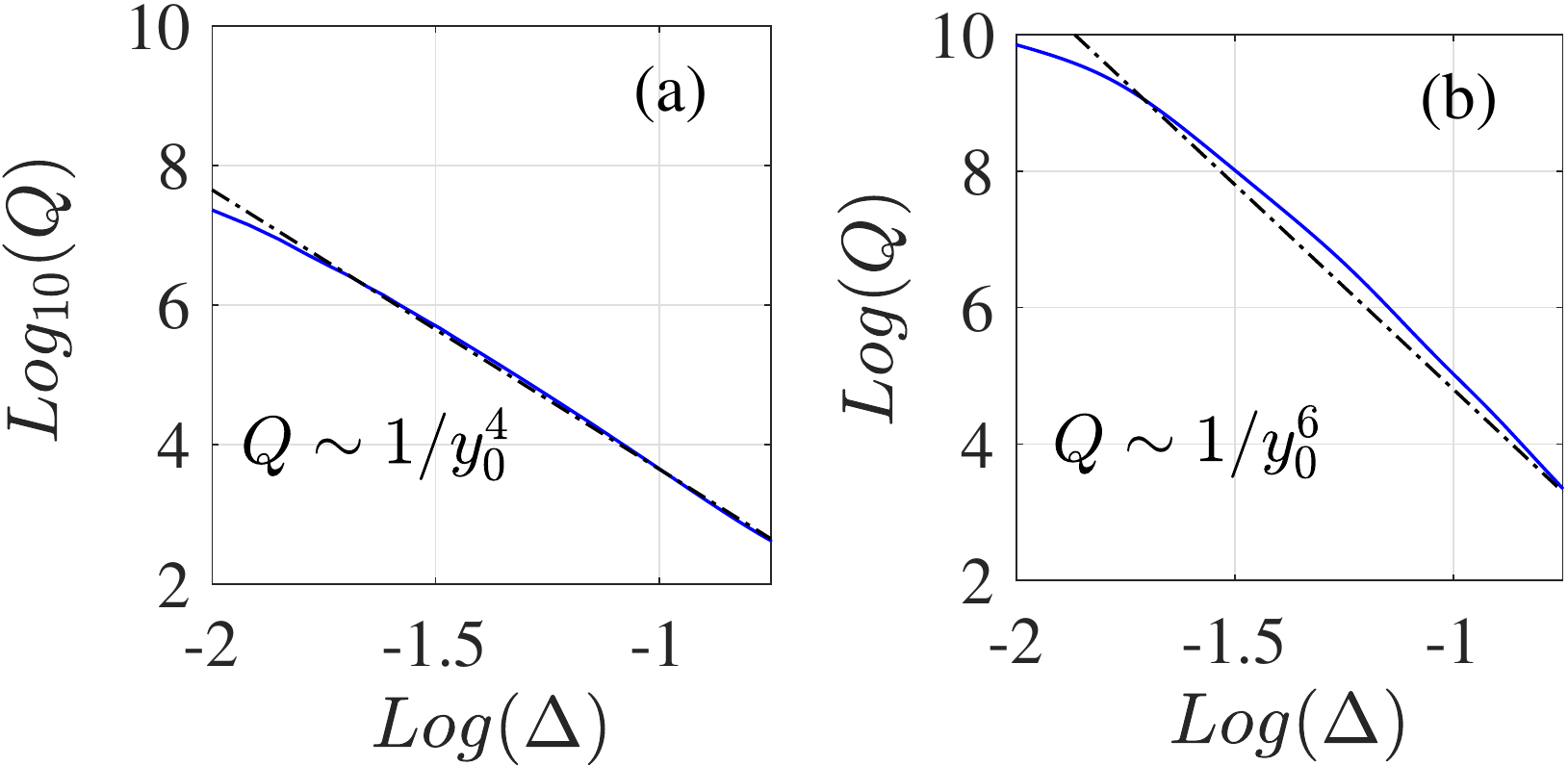}
\caption{ (Color online) Dependence of $Q$-factor vs position of rod in waveguide
in Log Log scale at the points of merging of topologically
protected BICs. (a) The case of circular rod shown in Fig.
\ref{fig11} and (b) the case of rectangular rod with cross-section
$a=0.3, b=b_c=0.911$ shown in Fig. \ref{fig11}. }
\label{fig12}
\end{figure}

These two TP BICs with $m=\pm 1$ are not distinguishable because
are related by the inverse $y \rightarrow -y$. However Fig. \ref{fig12}
(a) brightly demonstrates effect of annihilation of two TP BICs
which merge into the SP BIC with zero winding number when rod with
the critical radius $R_c=0.1656$ takes the symmetrical position
$\Delta\rightarrow 0$. Owing to Log Log scale of dependence we
obtain that $Q\sim \frac{1}{\Delta^4}$. If the radius of rod were
different from the critical one we would have standard quadratic behavior
$Q\sim \frac{1}{\Delta^2}$. First, these phenomenon
was demonstrated in periodical array of cylindrical rods in which
two off-$\Gamma$ BICs with winding numbers $m=\pm 1$ were merged
into the SP BIC at $\Gamma$ point with $m=0$
\cite{B&M2017}. This phenomenon is similar to the case of merging
of BICs observed in the photonic system, where topological
charges move toward  the $\Gamma$-point in first Brillouin zone at
momentum space \cite{Zhen,B&MPRL,Yuan2017,Jin2019,Zeng2021a}. It is
important to note that the merging of BICs does not mean existence
of two BICs at the same point of parametric space, i.e.,
degeneracy of BICs. In fact, for approaching to the merging point
accidental BICs vanish compared to the SP BIC.

\begin{figure}[ht!]
\centering
\includegraphics[width=0.5\linewidth]{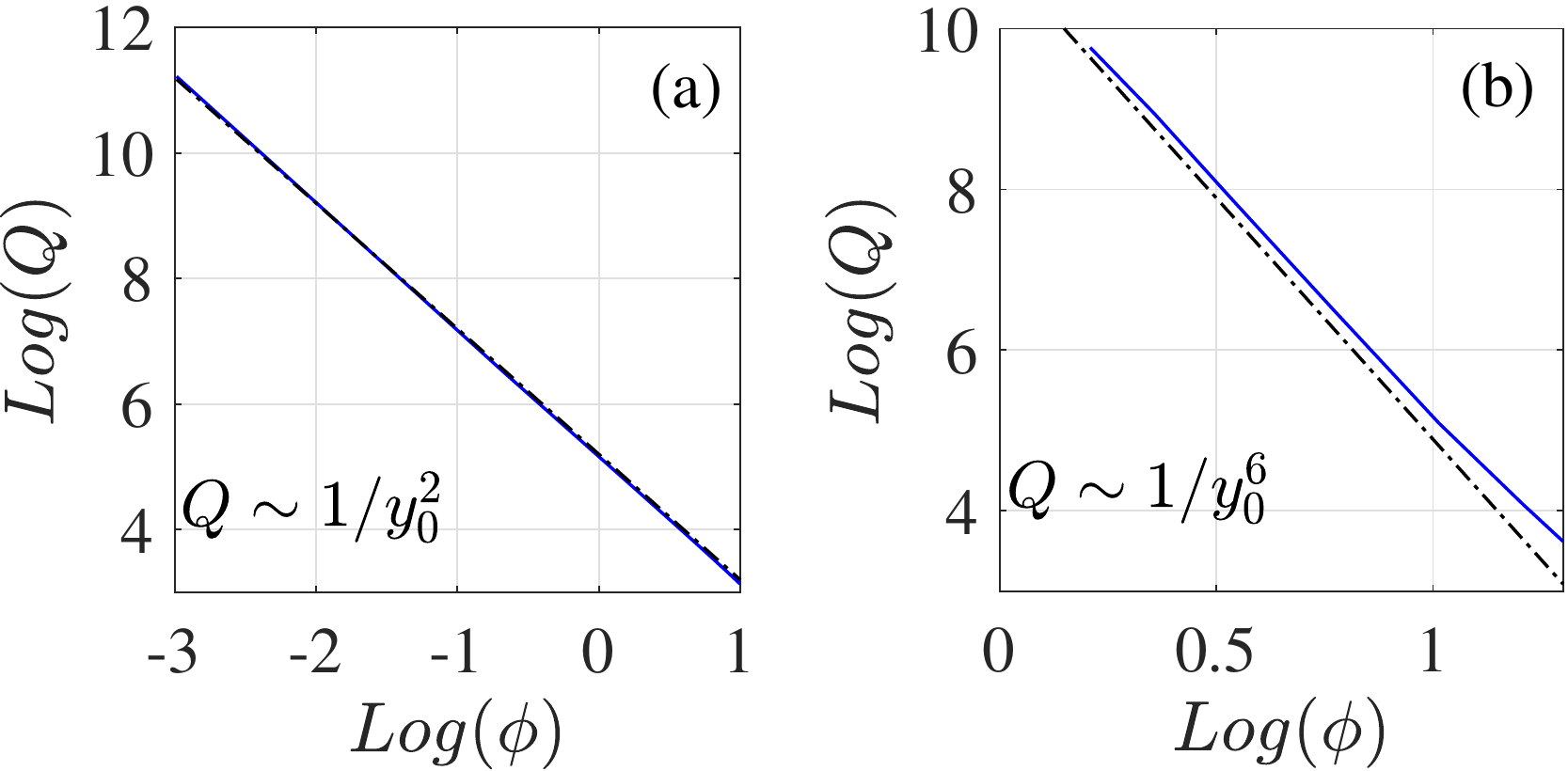}
\caption{ (Color online) Half width of resonances vs shift of
rectangular rod $a\times b$ with $n = 2.05$ and length $a$ at
$b=0.3$. Insets show patterns of BICs ($E_z$ of electric field). }
\label{fig13}
\end{figure}

Fig. \ref{fig13} demonstrates similar effects for the case of
rectangular rod $a \times b$ in two-parametric space of its
length $a$ and position $\Delta$.
At the merging point of two TP BICs with SP BIC the behavior of $Q$-factor
turns into $Q\sim \frac{1}{\Delta^6}$ as shown in Fig. \ref{fig12} (b).

\begin{figure}[ht!]
\centering
\includegraphics[width=0.5\linewidth]{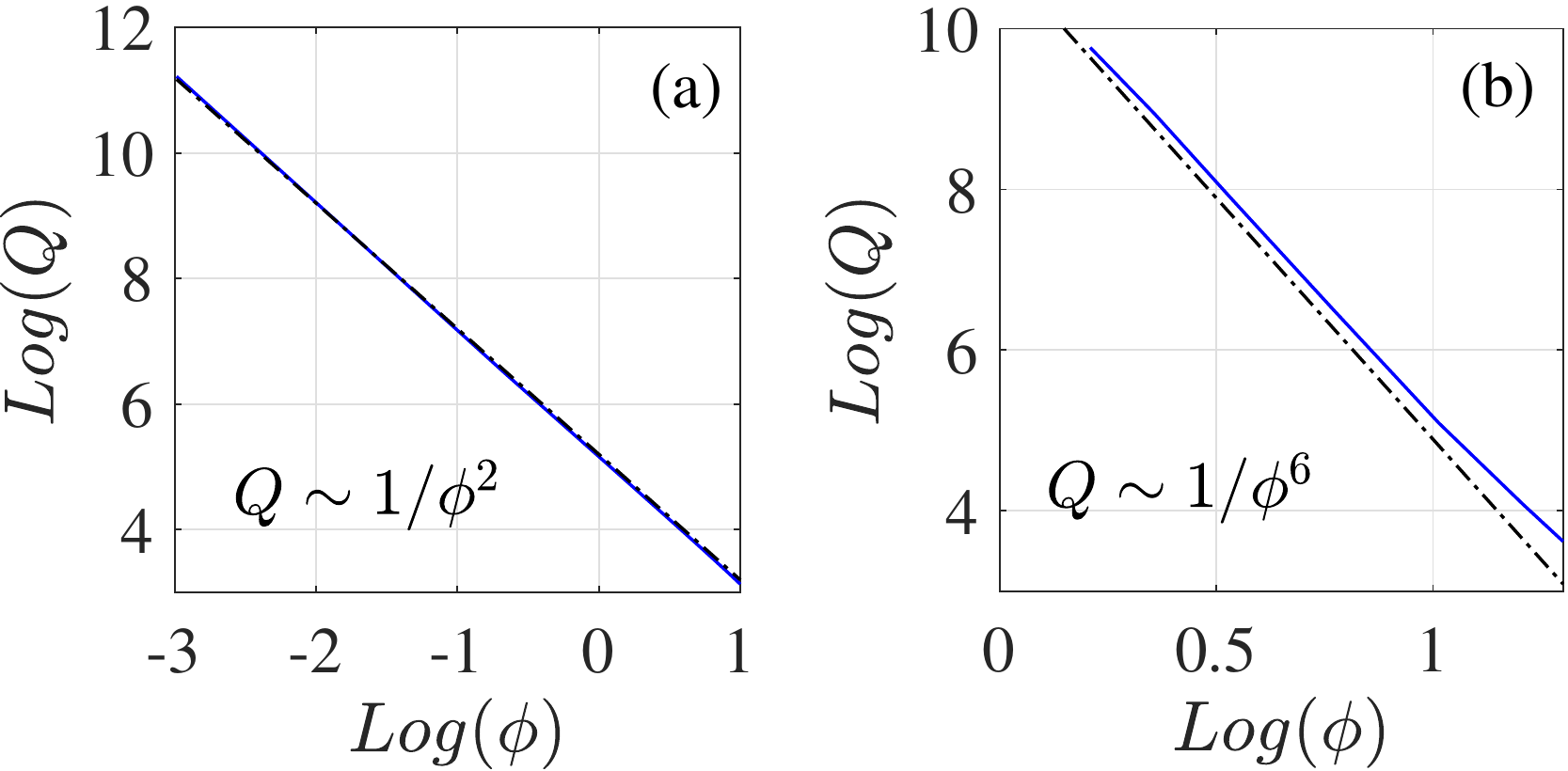}
\caption{ (Color online) Dependence of $Q$-factor vs rotation angle of two rectangular
rods with $a=0.6, b=0.3$
positioned symmetrically in waveguide in Log Log scale at the
points  (a) beyond  merging of topologically protected BICs marked
by star in Fig. \ref{fig8} (b) at $L=0.9$  and (b) at the point
of merging marked by open green circle in Fig. \ref{fig8} (b) at
$L=0.989$.} \label{fig14}
\end{figure}

The case of two rectangular rods brings a novelty of the merging of three
BICs, two Fabry-Perot BICs with winding numbers $m =\pm 1$ and
one SP BIC with $m=0$. Respectively, beyond the point of
merging (marked by star in Fig. \ref{fig8} (b)) we have standard
quadratic behavior of the $Q$-factor as demonstrated in Fig.
\ref{fig14} (a). However, at the points of merging marked by open circles in \ref{fig8} (b) we obtain
strong dependence $Q\sim \frac{1}{\Delta^6}$ as Fig. \ref{fig14} (b) shows.
\begin{figure}[ht!]
\centering
\includegraphics[width=0.45\linewidth]{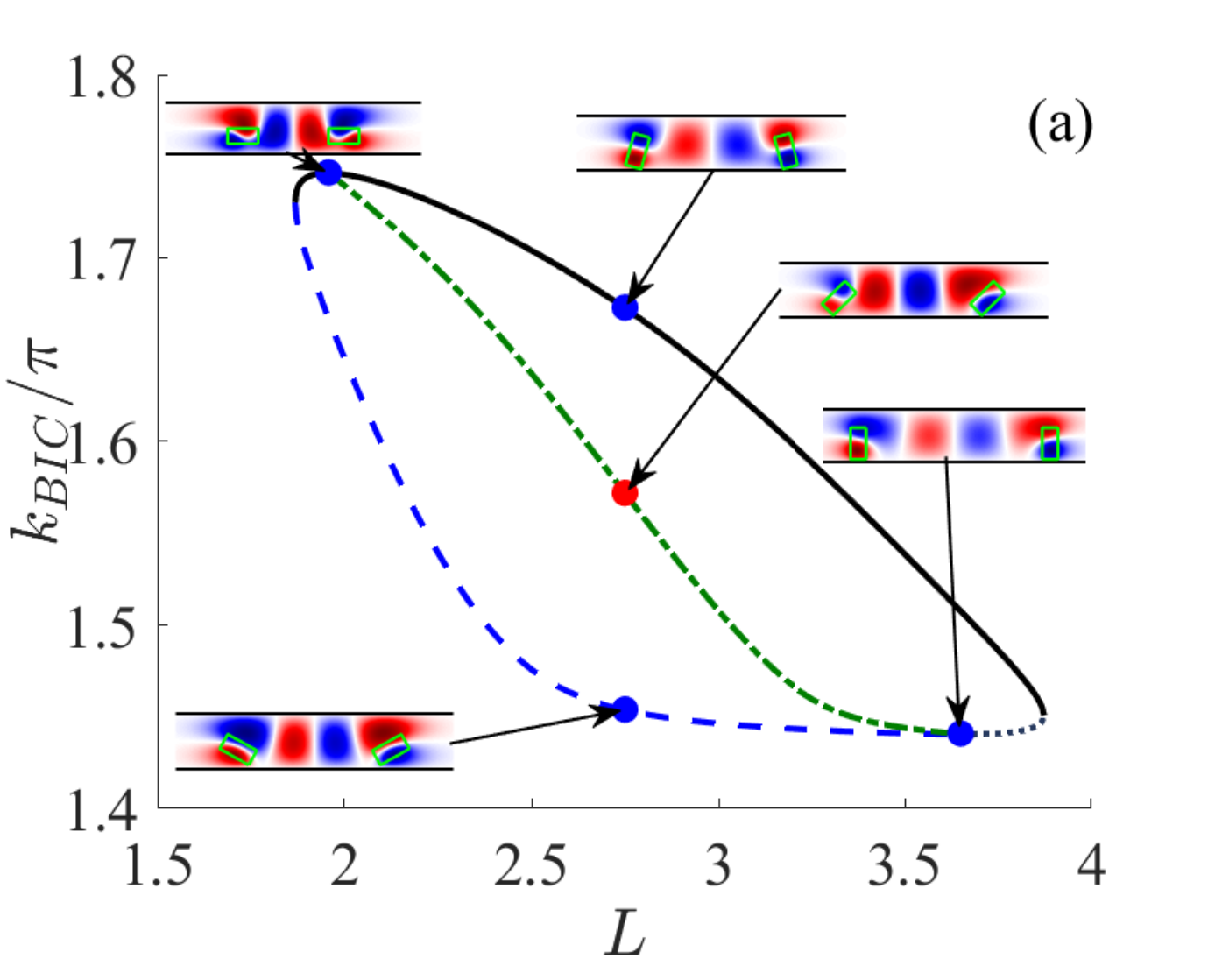}
\includegraphics[width=0.45\linewidth]{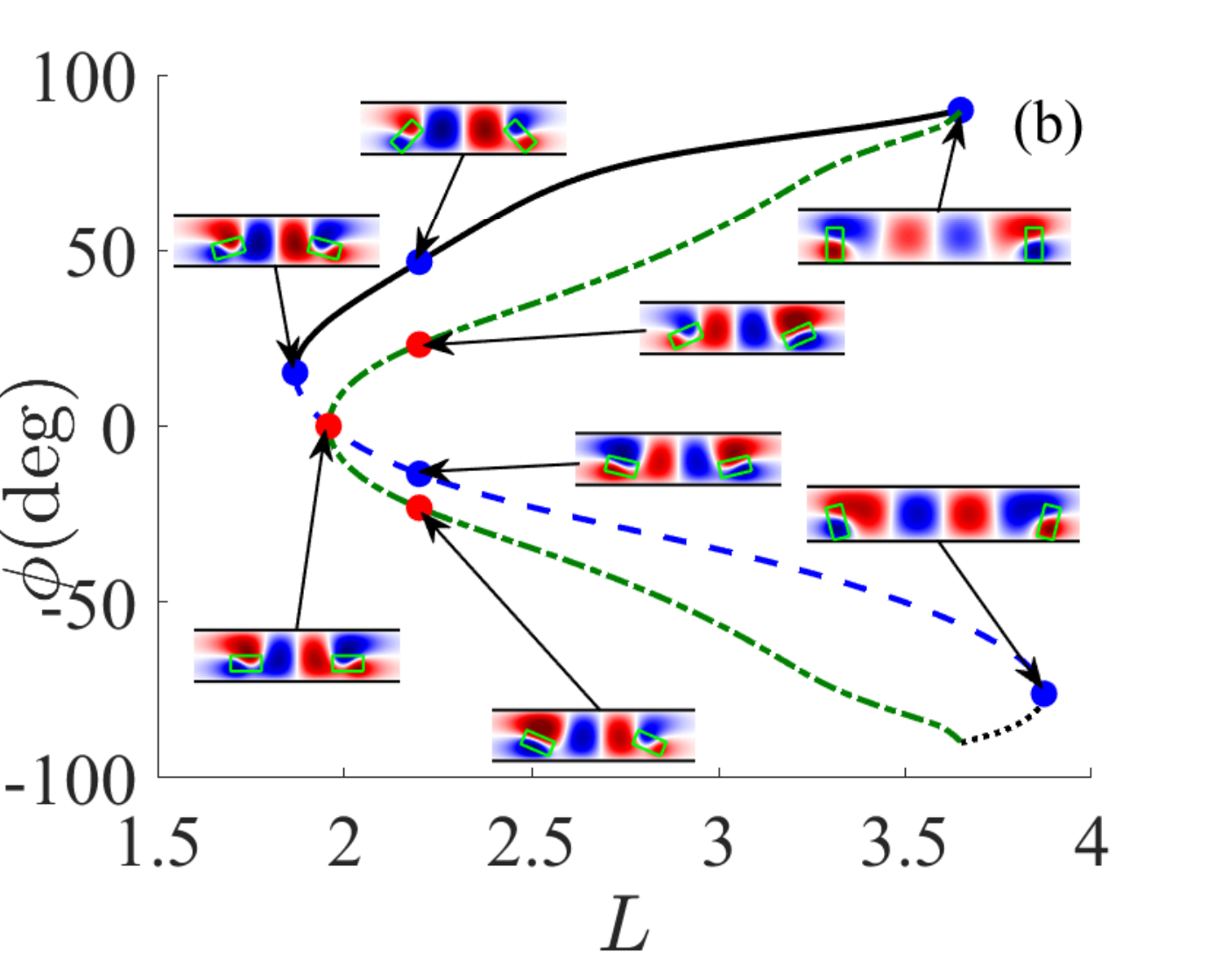}
\caption{ (Color online) Curves of existence of the Fabry-Perot BICs vs (a) distance
between rectangular rods and frequency and (b) angle of rotation
and distance between rods at $\Delta=1/2$. Dash-dotted
line show curves of Fabry-Perot SP BICs at $\phi=0, 90^{\circ}$. Green
solid line and dash gray lines show topologically protected
in-phase BICs and anti-phase BICs, respectively. Points of
BICs merging are marked by open circles. }
\label{fig15}
\end{figure}
Crossover the degree of enhancement of the $Q$-factor is related
to bifurcation of imaginary parts of TP BICs as Fig. \ref{fig11}
and \ref{fig13} show. That result follows from the algebra of
bifurcation \cite{Yuan&Lu2017,B&M2017,Jin2019,Zeng2021,Huang2022}.

One can see in Fig. \ref{fig8} points in which SP BICs at $\phi=0,
90^o$ (dash-dot blue lines) coalesce with Fabry-Perot BICs
evolving with angle of in phase or anti phase rotation of rods
shown by solid green and dash gray lines. These Fabry-Perot BICs
have morphology cardinally different from the SP BICs as seen from
insets in Fig. \ref{fig8}. For each angle $\phi\neq 0$ the
Fabry-Perot BIC has to be tuned by distance between rods $L$
making an analogy with accidental BICs. Fig. \ref{fig15}
demonstrates rich variety of merging effects of BICs with
different winding numbers in the three-parametric space of
frequency, distance between the rectangular rods and rotation
angle.

\section{Conclusions and discussion}

A simple system of plane waveguide consisted of two parallel metal
planes with integrated one or two dielectric quartz or silicon
rods demonstrates abundance of various BICs classified as symmetry
protected (SP), accidental, Friedrich-Wintgen and Fabry-Perot.
Moreover, we show numerous points of merging of different
topologically protected BICs in two-parametric space. These events
give rise to change the power degree in asymptotic behavior of the
$Q$-factor that has principal importance for numerous applications
of BICs. First, this phenomenon of BICs merging was observed in
the photonic system, where topological charges move toward the
$\Gamma$-point in first Brillouin zone at momentum space
\cite{Zhen,B&MPRL,Yuan2017,Jin2019,Zeng2021a}. Quite recently the
two-parametric space was expanded onto the frequency and coupling
strength of two acoustic resonators \cite{Huang2022}. In the
present paper we go further by introducing parameters determining
position of rods, distance between them, orientation of
rectangular shaped rods and frequency. It is important that all
parameters can be easily varied experimentally in comparison to
the case of topologically protected BICs in momentum space in the
one- and two-dimensional periodical arrays of dielectric
particles. It is clear that one can consider three-parametric
space in which we can observe the lines of TP BICs.

Finally, we discuss the extend to which extend the present
system of plane metallic waveguide with integrated dielectric rod
is preferable compared to periodic array of rods with account of
material losses. The comparison crucially depends on the frequency
range in which BICs are supposed to be observed. For example, in
GHz range the $Q$ factor of SP BICs in the GHz range in the array
of ceramic disks is saturated by a value 4000 \cite{Sadrieva2019}.
While the array of Si disks with $\tan=4\cdot 10^{-4}$ we have
$Q_{max}=2500$. In this range the silver surface can be considered
as perfectly conducting material, and therefore the main
contribution is due to material losses of rod integrated into
waveguide. Fig. \ref{fig16} presents results of  COMSOL
Multiphysics computations which show the frequency of SP BICs and
its $Q$-factor versus the radius of rod that shows considerable
gain of the present system compared to the periodical array of
rods in the GHz range.
\begin{figure}[ht!]
\centering
\includegraphics[width=0.5\linewidth]{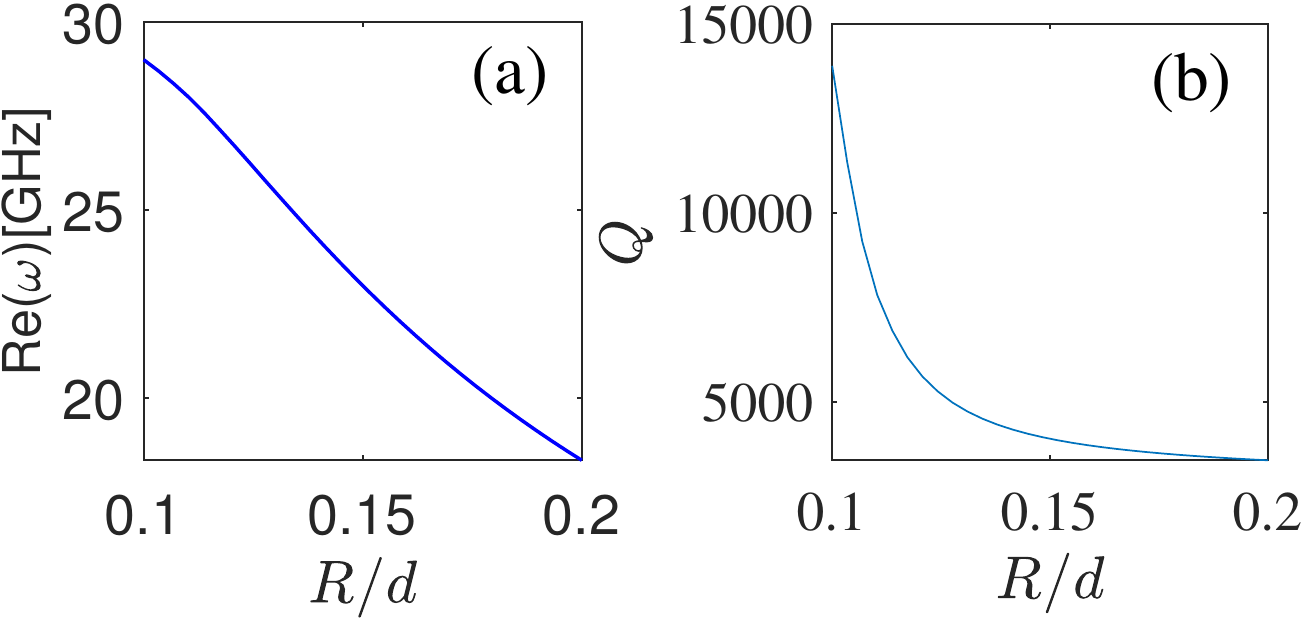}
\caption{ (Color online) Dependence of frequency (a) and the
$Q$-factor (b) of the SP quasi BIC vs radius of Si rod in the GHz
range with account of material losses of rod.} \label{fig15}
\end{figure}
In the optical range the main adverse factor is the surface
impedance of silver metal that considerably restricts the
$Q$-factor as shown in Fig. \ref{fig17} by use of data of Ref.
\cite{Johnson1972}. Nevertheless in red line range the $Q$-factor
of the SP BIC can exceed a value 2000.
\begin{figure}[ht!]
\centering
\includegraphics[width=0.5\linewidth]{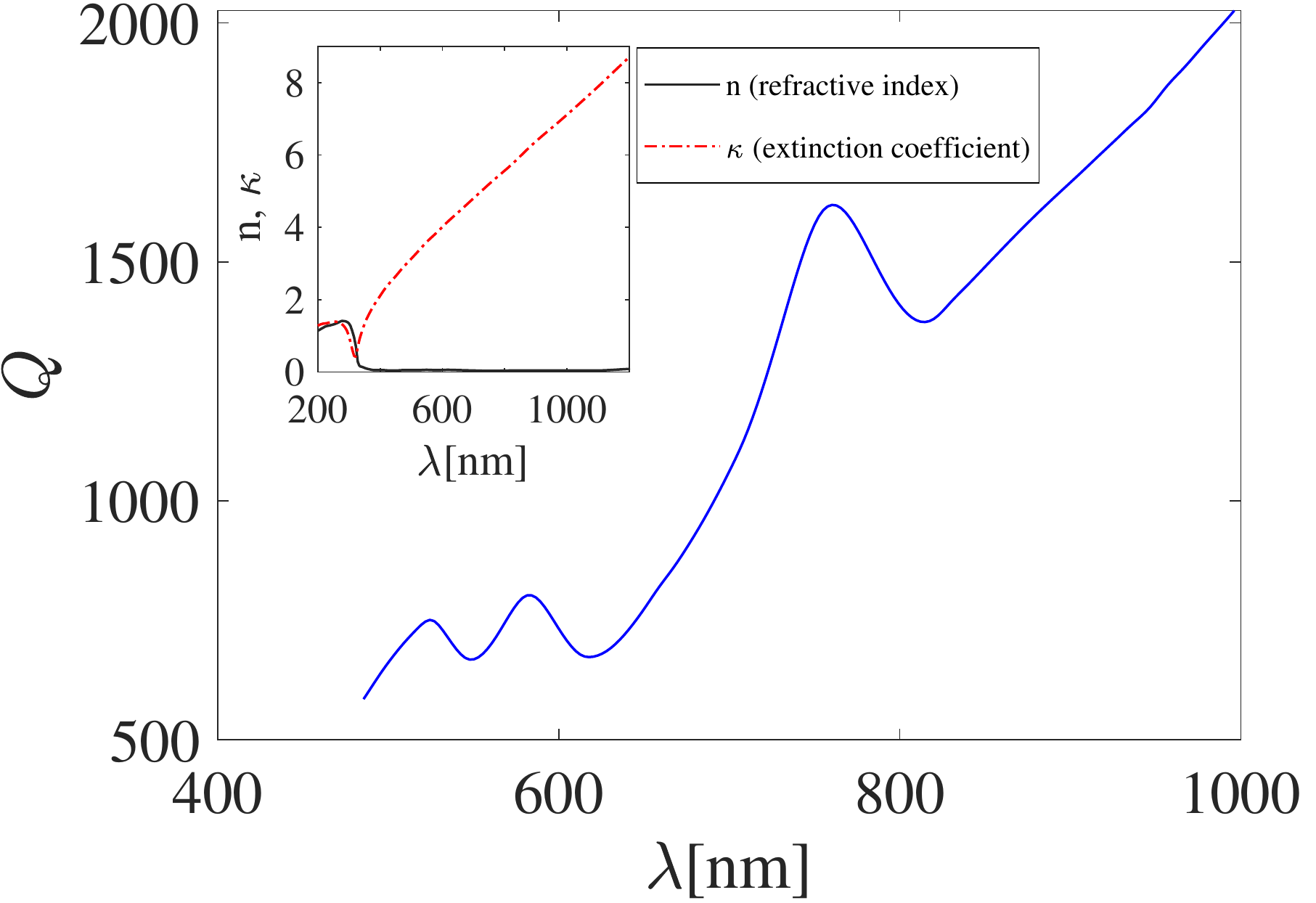}
\caption{ (Color online) Dependence of $Q$-factor of the SP quasi
BIC vs wavelength for the rod integrated into silver waveguide in
the optical range with account of surface impedance of silver.}
\label{fig17}
\end{figure}

\section{Acknowledgements}
We are grateful to Lujun Huang, Andrey Miroshnichenko and Yi Xu
for discussions. The research was supported by Russian Science
Foundation with Grant number 22-12-00070.
%
\end{document}